# Atmospheric Supply of Hydrogen Cyanide Is Not the Rate-limiting Step for Prebiotic Chemistry across Rocky Exoplanets

Gergely Friss[1,2], Paul I. Palmer[2,3], Marrick Braam[4], and Ken Rice[2,5]
[1] School of Physics and Astronomy, The University of Edinburgh, Edinburgh, EH9 3FF, UK; gergely.friss@ed.ac.uk
[2] Centre for Exoplanet Science, The University of Edinburgh, Edinburgh, EH9 3FF, UK; pip@ed.ac.uk
[3] School of GeoSciences, The University of Edinburgh, Edinburgh, EH9 3FF, UK
[4] Center for Space and Habitability (CSH), University of Bern, Gesellschaftsstrasse 6 (G6), CH-3012, Bern, Switzerland
[5] SUPA, Institute for Astronomy, University of Edinburgh, The Royal Observatory, Blackford Hill, Edinburgh, EH9 3HJ, UK

## Abstract

Hydrogen cyanide (HCN) is crucial for the RNA World hypothesis, forming biomolecules essential for early life. Life likely emerged around 4 billion yr ago during the early Archean Eon, a period on Earth with a fainter Sun, frequent impacts, and a weakly reducing atmosphere. Warm little ponds (WLPs) are hypothetical protective aqueous environments that help explain the emergence and evolution of fragile prebiotic chemistry in such a hostile environment. WLPs need to undergo cycles of evaporation and rehydration, concentrating prebiotic molecules that increase the likelihood of (de-)polymerisation and forming early RNA molecules. We use a 1D model of atmospheric chemistry to compare atmospheric HCN delivery to WLPs with exogenous sources. Using early Archean Earth as our baseline, we examine the sensitivity of atmospheric HCN delivery to the atmospheric C/O ratio, semimajor axis, assumed stellar host type, and methane budget, exploring conditions across rocky exoplanets. We find that atmospheric HCN delivery is sensitive to these parameters, but its values generally exceed those of meteoritic delivery and our baseline Archean Earth. Planetary atmospheres with higher C/O ratios within the habitable zones of G stars and those closely orbiting M-dwarfs deliver the most atmospheric HCN. We find that atmospheric HCN delivery is remarkably robust, so this molecule is likely not the rate-limiting step for the emergence of prebiotic chemistry on rocky exoplanets. This finding, with important caveats, potentially increases the probability of life emerging on other worlds.

## 1. Introduction

Microfossil evidence, supported by geochemical analysis, suggests that life emerged relatively early in Earth's history, potentially as far back as 3.5–3.8 billion yr ago in the early Archean eon (A. P. Nutman et al. 2016; T. Djokic et al. 2017, 2021) or even earlier in the late Hadean eon (E. A. Bell et al. 2015). Interpretations of the earliest signs of life on Earth are based on isotopic evidence, although this method is not without controversy (E. A. Bell et al. 2015; A. P. Nutman et al. 2016). Both lines of evidence represent a top-down perspective that tells us nothing about the origin of the fossilized organic structures or provides evidence of any anticipated follow-on biological processes, such as cell division. A complementary bottom-up approach is to develop prebiotic chemical networks that help to explain the organic structures inferred from the microfossils. Ultimately, these networks need to use molecules that we think were ubiquitous in the Archean environment as feedstock to develop long-chain biomolecules, including nucleobases, amino acids, carboxylic acids, sugars, and other complex molecules. Our knowledge of these networks is incomplete, requiring substantial empirical data to evaluate individual reactions and to bridge laboratory data and the less-controlled real-life environments.

Based on collective paleo records and associated model interpretation, the Archean atmosphere was likely characterized by $N_2$ and $CO_2$ as bulk species, CO, $H_2$, and methane as trace species, and oxygen in extremely low amounts (D. C. Catling & K. J. Zahnle 2020). Inconclusive evidence suggests that even in the preceding Hadean eon, continents and oceans began to form ( E. A. Bell et al. 2015; T. M. Harrison et al. 2017). The Archean (and Hadean) atmosphere was likely influenced by meteoritic bombardment ( B. K. D. Pearce et al. 2017; D. C. Catling & K. J. Zahnle 2020). In which case, molten iron from impacted meteors would have reacted with surface water to release molecular hydrogen into the atmosphere, reinforcing the influence of outgassing (C. Brachmann et al. 2025; J. P. Itcovitz et al. 2022). As such, the early Archean atmosphere would likely have been reducing (at least locally) so that it would have been easier to form carbon–hydrogen and carbon–carbon bonds with prebiotic chemistry, which form the backbone of organic molecules (D. C. Catling & K. J. Zahnle 2020; K. J. Zahnle et al. 2020; N. F. Wogan et al. 2023).

Hydrogen cyanide (HCN) is a particularly attractive molecule in prebiotic chemistry due to its chemical properties and plausible abundance on the early Earth (F. Tian et al. 2011; P. B. Rimmer & S. Rugheimer 2019; B. K. D. Pearce et al. 2022). HCN can form in the atmosphere through high-energy events such as lightning and UV photochemistry (P. B. Rimmer & S. Rugheimer 2019; B. K. D. Pearce et al. 2022) or meteoritic impacts (R. Saladino et al. 2018), which provide the energy needed to break the strong triple bond of nitrogen gas ($N_2$) and allow it to combine with carbon. It is also a soluble compound, so that it can be rained out of the atmosphere as an efficient mechanism of delivery to the planetary surface. The carbon–nitrogen triple bond in HCN is a highly reactive functional group, making it a potential precursor for a wide range of essential biomolecules once on the planetary surface: HCN chemistry can lead to the formation of, for example, various amino acids via the Strecker synthesis (A. Strecker 1854), a process consistent with the abiotic amino acid synthesis demonstrated in experiments, e.g., the Miller-Urey experiment (S. L. Miller 1953). HCN in an aqueous solution can polymerize into more complex organic species, such as sugars, lipids, nucleobases and even fragments of RNA chains, a process catalyzed by the presence of phosphate (D. Ritson & J. D. Sutherland 2012) and a repeating wet–dry cycle that concentrates dissolved organic molecules, increasing the probability of polymerization of HCN (B. H. Patel et al. 2015; B. Damer & D. Deamer 2020; B. K. D. Pearce et al. 2022). Furthermore, HCN could have played a vital role in increasing the solubility of phosphate minerals, a requirement for building complex molecules of RNA and DNA (B. Burcar et al. 2019).

Warm little ponds (WLPs) on the surface of early Earth have been proposed as an ideal environment for such aqueous prebiotic chemistry. WLPs could have originated when early volcanic islands formed or due to meteoritic impacts (B. K. D. Pearce et al. 2017; B. Damer & D. Deamer 2020). Although we can only speculate on their size, such water bodies would have provided consistent warmth (N. H. Sleep 2010; B. K. D. Pearce et al. 2017; C. S. Cockell 2020), driven chemical reactions by circulating hot, mineral-rich fluids (B. H. Patel et al. 2015), and exposed diverse mineral catalysts (B. Burcar et al. 2019). Wet–dry cycles could have been driven by many environmental factors, such as periodic changes in volcanic geysers, tidal pools on coastal rocks (B. Damer & D. Deamer 2020), or seasonal climate (B. K. D. Pearce et al. 2017), that would have introduced a diverse range for their duration and periodicity. The wet–dry cycles, connectivity, and potential delivery of organics (both exogenous and atmospheric) make WLPs an ideal, protected, and energetically favorable setting for the complex prebiotic chemistry necessary for life to emerge (B. H. Patel et al. 2015; B. K. D. Pearce et al. 2017; B. Damer & D. Deamer 2020).

The exploration of early Earth is also relevant to our understanding of rocky exoplanets and the possible surface environments they could support. A. Bohl et al. (2026) reported 67 rocky planets (out of 294 according to data from the NASA Exoplanet Archive (J. L. Christiansen et al. 2025) obtained using the same search criteria) that orbit in the empirical habitable zone (HZ) of their respective host star, meaning that liquid water could exist on their surface (J. F. Kasting et al. 1993; R. K. Kopparapu et al. 2013). Rocky exoplanets in the HZ are easier to discover using current technologies when they are orbiting M stars. There are two implications relevant to our work, with regard to the habitability of planets orbiting such stars that are smaller and cooler than the Sun. First, the HZ is much narrower and closer to the star than it is in the solar system (R. K. Kopparapu et al. 2013). Thus, exoplanets have closer-in orbits and are typically tidally locked in a 1:1 spin–orbit resonance (SOR) with permanent day and night sides on the planet, although higher SORs are possible as well (R. Barnes 2017, and references therein). Studies have shown that periodic changes in water vapor can be introduced by oscillating atmospheric waves on planets with a 1:1 SOR (M. Cohen et al. 2022) and by the varying stellar irradiation received by a planet in a 3:2 SOR (M. Braam et al. 2025). Second, these cooler stars have a radiation spectrum that is shifted toward longer wavelengths (e.g., A. L. Shields et al. 2016). Such orbital configurations will introduce different wet and dry cycles needed by WLPs, provided by axial tilt on Earth, to process the prebiotic molecules and a different chemical environment in the atmosphere. Understanding the differences between our baseline calculations for early Earth, which we know led to the emergence of life, and calculations for a range of rocky exoplanet environments will provide insights into the ability of these planets to potentially support prebiotic chemistry.

Here, we explore the physical parameter space, encapsulating a range of rocky exoplanet environments, to examine the sensitivity of the atmospheric chemistry and subsequent surface delivery of HCN as initial steps of prebiotic chemistry. We demonstrate that, albeit with important and acknowledged caveats, HCN formation and delivery to WLPs are robust across a wide range of planetary conditions and are likely not the rate-limiting steps for the emergence of life on rocky exoplanets.

## 2. Methods

We use the open-source VULCAN (S.-M. Tsai et al. 2017, 2021) 1D steady-state photochemical model to explore the chemistry and wet deposition of atmospheric HCN on Archean Earth, and to examine the sensitivity of these calculations to changes in planetary-scale parameters that encapsulate the environments on a diverse set of possible rocky exoplanets. We describe the fundamental equations of the model and the chemical network we use in Appendix A and refer the reader to the original papers for a more detailed description of the code. Here, we describe our development of implementing wet deposition of trace gases and our numerical experiments.

### 2.1. Wet Deposition of Trace Gases

As 1D modeling of an atmosphere cannot include global atmospheric circulation and precipitation (F. Giorgi & W. L. Chameides 1985), we implement a first-order, time-independent parameterization for water rain, which we combine with Henry's law of solubility to obtain the rainout rate of HCN through in-cloud scavenging. We calculate the rainout rate of water ($k_{\mathrm{H_2O}}$, units of $\mathrm{s^{-1}}$) using eddy diffusion ($K_{zz}$, units of $\mathrm{cm^2\,s^{-1}}$) and gas phase water abundance ($n_{\mathrm{H_2O}}$, units of $\mathrm{cm^{-3}}$) as described in F. Giorgi & W. L. Chameides (1985):

$$k_{\mathrm{H_2O}}(z) = -\frac{1}{n_{\mathrm{H_2O}}(z)}\frac{d}{dz}\Phi_{H_2O}(z), \quad (1)$$

where $\Phi_{\mathrm{H_2O}}(z) = -K_{zz}(z)n_M(z)d(X_{H_2O}(z))/dz$ is the vertical flux of water with $n_M$ and $X_{\mathrm{H_2O}}$ being the total number density of the atmosphere (units of $\mathrm{cm^{-3}}$) and mixing ratio of water (unitless), respectively, and $z$ is the height in centimeters. We treat rain as a new chemical species, similarly to how VULCAN treats condensates as chemical species, which are produced but not destroyed and are not considered part of the gas phase. This approach ensures numerical stability. We use the water rain rate to calculate in-cloud scavenging rate of species $i$ (units of $s^{-1}$) as follows

$$k_{i,\mathrm{scav}} = r_i f_i k_{\mathrm{H2O}}, \tag{2}$$

where $r_i$ is the retention factor (unitless) and $f_i = K_i^* LRT/(1 + K_i^* LRT)$ is the equilibrium fractionation (unitless), i.e., what fraction of the molecules of species $i$ gets dissolved in the raindrops, with $K_i^*$ being the effective Henry's law constant (units of M/atm), $R$ is the universal gas constant (units of M/atm/K) and $L$ is the fraction of condensed water with regards to the gas content of the given cell (D. J. Jacob et al. 2000; G. Luo et al. 2020). The retention factor is taken to be unity for $T > 268$ K and 0.02 otherwise, as a way to force a difference between liquid and solid water.

All scripts used in this study for code development, along with output data and code for analyzing the data, are available on GitHub[6] under an MIT License, and version 1.1 is archived on Zenodo: DOI:10.5281/ZENODO.18633433.

### 2.2. Model Parameters and Constructing Numerical Experiments

Before exploring the chemical effects of the physical parameter space, we establish a fiducial model of the Archean Earth, ~4 Gyr ago. Then, we explore how the carbon-to-oxygen (C/O) ratio, semimajor axis, type of host star, initial methane abundance, and meteoritic bombardment affect the atmospheric chemistry by only changing one of these parameters at a time while keeping others the same (when reasonable). Last, we study the conjoint effects on HCN deposition of the C/O ratio, the semimajor axis, and the type of host star.

#### 2.2.1. Archean Earth Baseline Calculation

We set the atmospheric pressure range to 1–5 × $10^{-8}$ bar and divide the atmosphere into 120 vertical layers. We use a general Earth eddy-diffusion profile (S. T. Massie & D. M. Hunten 1981) that is already included in VULCAN (S.-M. Tsai et al. 2021) and is used in several other studies (F. Tian et al. 2011; R. Hu et al. 2012; S. Ranjan et al. 2020; B. K. D. Pearce et al. 2022). We impose a diurnal cycle, we set the stellar zenith angle to 58°, which is chosen to represent the global mean incoming stellar radiation, and we fix the surface albedo at 0.105 to represent an Archean Earth covered with oceans and land, with the latter being made of dark basalt (E. T. Wolf & O. B. Toon 2014). We implement the surface boundary conditions from B. K. D. Pearce et al. (2022) that incorporate, among others, volcanic degassing, evaporation of surface water, and HCN production from lightning chemistry (Table 1) and set the top of the atmosphere boundary conditions to zero so there is no atmospheric escape of H and He.

The initial vertical mixing ratio of water is set to decrease linearly from 0.01 at the surface to $10^{-6}$ at $p = 0.14$ bar, above which it remains constant. This vertical structure is almost the same as described in B. K. D. Pearce et al. (2022), except for a more realistic mixing ratio above the tropopause. Following D. C. Catling & K. J. Zahnle (2020), our initial atmosphere also consists of $CO_2$, $CH_4$, and $O_2$ with constant mixing ratios of 0.1, 5 × $10^{-3}$, and $10^{-7}$, respectively, while the main constituent is $N_2$. We account for the Sun being ~30% less luminous than today (J. N. Bahcall et al. 2001; M. W. Claire et al. 2012; D. C. Catling & K. J. Zahnle 2020) by implementing an early stellar flux 4 Gyr ago (a 0.6 Gyr old Sun) using the parameterization from M. W. Claire et al. (2012).

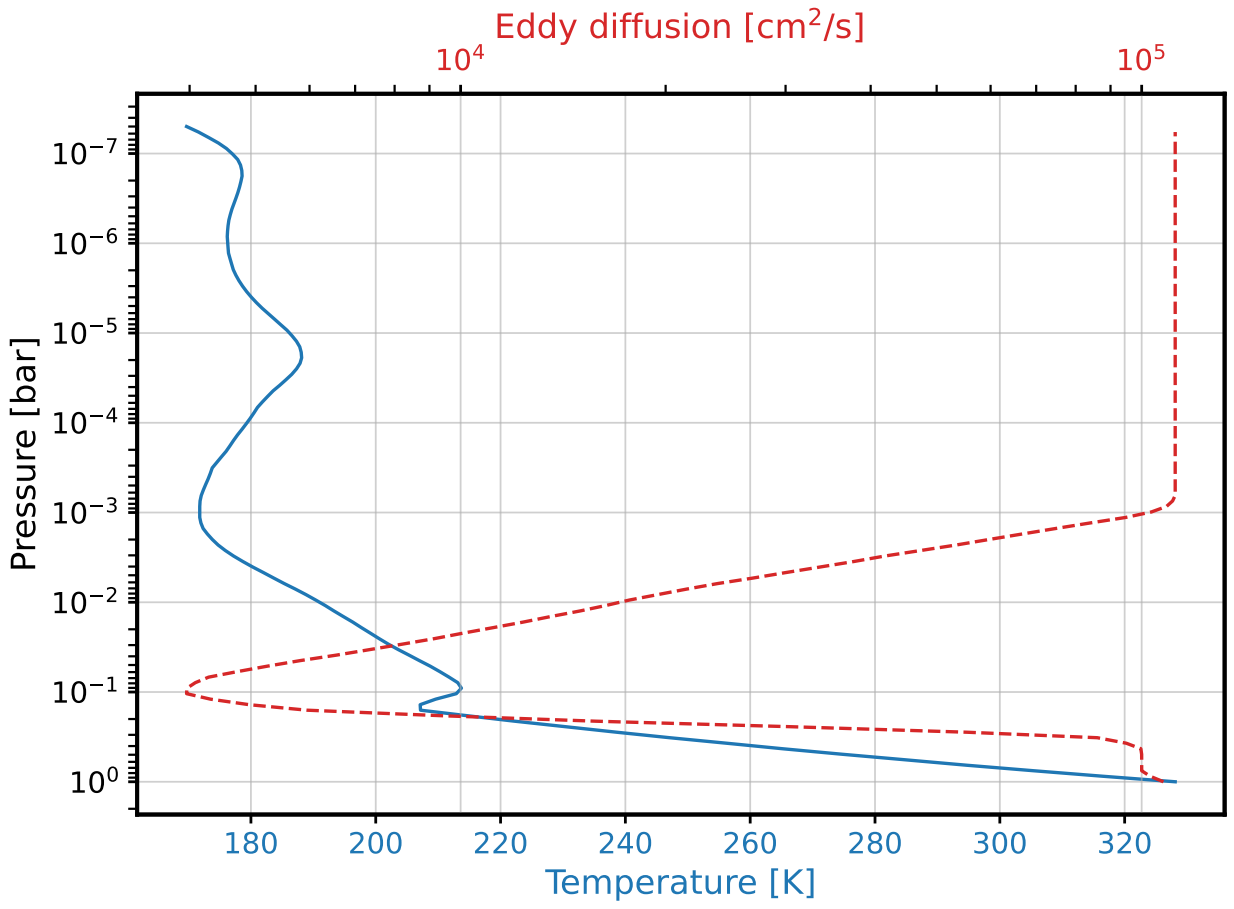

**Figure 1.** Temperature–pressure and eddy-diffusion profile in our base Archean simulation. The eddy-diffusion profile is Earth-like and fixed throughout all our models.

**Table 1**
Surface Boundary Conditions Used in Our Archean Simulation Taken from B. K. D. Pearce et al. (2022)

| Species | Flux ($cm^{-2}$ $s^{-1}$) |
|---|---|
| $H_2$ | $2.3 \times 10^{10}$ |
| $CO_2$ | $3.0 \times 10^{11}$ |
| $CH_4$ | $6.8 \times 10^{8}$ |
| $H_2O$ | $2.0 \times 10^{9}$ |
| H | $9.7 \times 10^{1}$ |
| CO | $1.8 \times 10^{5}$ |
| OH | $3.5 \times 10^{3}$ |
| NO | $7.4 \times 10^{3}$ |
| HCN | $2.9 \times 10^{-6}$ |

Given the initial composition of the atmosphere, stellar irradiation profile, and planetary parameters, such as semimajor axis (1 au) and size (1 $R_\oplus$), we calculate the temperature–pressure ($T$–$P$) profile of our model Archean Earth. For that purpose, we use the open-source HELIOS 1D radiative transfer model (M. Malik et al. 2017, 2019b) that is capable of calculating $T$–$P$ profiles in radiative–convective equilibrium for both gaseous and rocky planets (M. Malik et al. 2019a; E. A. Whittaker et al. 2022). For all our calculations, we set the equilibrium temperature of the planet to 43.3 K (B. K. D. Pearce et al. 2022) and assume a cloudless atmosphere with efficient day–night heat transport. We present the $T$–$P$ profile of our Archean Earth model in Figure 1.

#### 2.2.2. Individual Parameter Studies

Prebiotic chemistry favors a more reducing atmosphere (P. B. Rimmer & S. Rugheimer 2019), and so we study how HCN chemistry is affected by varying the initial atmospheric composition of our test planet. To achieve that, we substitute a

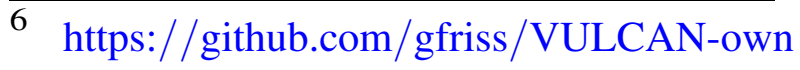

6 https://github.com/gfriss/VULCAN-own

given amount of $CO_2$ with CO in our initial atmosphere. Our choice of CO is based on the possible volcanic outgassing of this molecule on rocky exoplanets and the early Earth (Y. Watanabe & K. Ozaki 2024). We progressively substitute 0%–100% of $CO_2$ with CO in equal increments over 15 simulations, resulting in C/O ratios spanning from 0.514 to 0.998.

The surface habitability of a rocky planet is highly sensitive to its semimajor axis. The exact location of the planet determines the amount of incoming stellar radiation, thereby controlling the surface temperature and the entire *T–P* profile of its atmosphere. We note that the semimajor axis also drives the atmospheric circulation regimes, due to tidal locking (L. Carone et al. 2015; T. M. Merlis & T. Schneider 2010; J. Haqq-Misra et al. 2018), but this cannot be explicitly resolved in a 1D model, and therefore we omit such effects from our study. Instead, we explore the sensitivity of atmospheric HCN delivery to the surface of our Archean Earth by placing the planet at 15 different semimajor axes, ranging from 0.839 to 1.333 au in equal increments that keep surface temperatures habitable (i.e., allow for liquid water on the surface). We keep the default atmospheric initial composition and early solar irradiance profile, and use HELIOS to calculate the surface temperatures that range 273.5–372.4 K. We show the resulting *T–P* profiles in Figure 2(h).

Motivated by the inherent diversity of stars across the Universe, and acknowledging that current observational capabilities predominantly yield rocky exoplanets orbiting M-type stars, we have modeled an Archean Earth analog orbiting 13 distinct stars. This approach enables us to investigate the dependence of atmospheric chemistry on varying stellar irradiation profiles. We obtain 11 stellar spectra from the MUSCLES spectral survey: four from version 2.2 (K. France et al. 2016; R. O. P. Loyd et al. 2016; A. Youngblood et al. 2016), five from version 2.3 (D. J. Wilson et al. 2021), and two from version 2.4 (P. R. Behr et al. 2023). We also adapt the modern solar irradiation profile from C. A. Gueymard (2018). Spectra from the MEGA-MUSCLES survey (D. J. Wilson et al. 2021, version 2.3) have negative values that can introduce numerical instabilities in VULCAN. To address this, we smooth the spectra from all surveys by replacing values below 10 erg $cm^{-2}$ $s^{-1}$ $nm^{-1}$ with the value that is above this threshold and belongs to the nearest wavelength. All spectra are scaled so that they give the flux at the stellar surface for the VULCAN and HELIOS models and are compared in Figure 3, including the default spectrum of the early Sun. We calculate the *T–P* profiles (Figure 2(l)) using HELIOS for each spectrum because the stellar irradiation profile affects the structure and chemistry of the atmosphere, which motivates the simultaneous change of two parameters in this case. To keep everything as controlled as we can, we position our model planets around their host stars so that their surface temperatures are within $\sim$1 K of our baseline Archean Earth calculation. Table 2 shows a comprehensive overview of the stellar and orbital parameters.

The atmospheric composition of a young exoplanet after its magma ocean phase is the subject of ongoing research, with studies reaching different conclusions about the amount of $CH_4$ present. We adopted a baseline of 5000 ppm $CH_4$ for our baseline Archean Earth simulation, consistent with geological data summarized by D. C. Catling & K. J. Zahnle (2020). The authors indicate that achieving these elevated concentrations would require biogenic sources. Other studies using coupled ocean-atmosphere models (J. Krissansen-Totton et al. 2018; M. A. Thompson et al. 2022) or coupled mantle-atmosphere models (P. Liggins et al. 2022, 2023) also suggest that such an atmosphere, with a high amount of $CH_4$ and no CO, might only be plausible if biogenic sources of $CH_4$ are present. As such, abiotic $CH_4$ abundances may be as low as 1–10 ppm on the early Earth (B. K. D. Pearce et al. 2022), based on equilibrium thermodynamics and hydrothermal flux estimates. Contrary to these findings, other studies considering equilibrium chemistry models for surface temperatures $<$600 K (P. Woitke et al. 2021) or coupled mantle-atmosphere chemistry (C. Brachmann et al. 2025) found that $CO_2$, $CH_4$, and $H_2O$ could coexist with high abundances if only abiotic sources are present. This case is the most likely for planets that orbit G stars and have an intermediate reducing mantle (C. Brachmann et al. 2025). As such, the results of their early Earth model are in good agreement with the Archean atmospheric composition (D. C. Catling & K. J. Zahnle 2020). Discrepancies between these studies may arise from several factors, including the choice of numerical models, initial elemental abundances, surface temperatures, and the specific chemical frameworks employed, specifically whether they incorporate kinetics or photochemistry. Furthermore, the varying approaches to mantle-atmosphere coupling may also contribute to this divergence. Ultimately, these results underscore the likely diversity of exoplanetary environments, suggesting that research must look beyond the C/O ratio to consider the inherent uncertainty of atmospheric methane budgets. To address this point, we report results from an experiment that maintains a constant C/O ratio and $CO_2$ mixing ratio while varying the initial $CH_4$ concentration from 1 to 5000 ppm. To ensure the C/O ratio remains fixed across all simulations, CO is introduced as a balancing agent.

The early solar system was a chaotic place with bombardment of the early Earth by smaller impactors, or large ones like the Moon-forming impact (K. J. Zahnle et al. 2020). Other stellar systems could go through a similar phase of bombardment, although their duration and intensity would vary for different host stars and configurations (T. Lichtenberg & M. S. Clement 2022). Impactors with an iron core would reduce the surface water to hydrogen through the reaction $Fe+H_2O \rightarrow FeO+H_2$, creating a locally more reducing atmosphere, which would support prebiotic organic chemistry (H. Genda et al. 2017; P. B. Rimmer & S. Rugheimer 2019; K. J. Zahnle et al. 2020). Motivated by the uncertainty on the meteoritic bombardment rate experienced by the Archean Earth, and what might be experienced by rocky exoplanets, and its potential involvement with enhanced HCN production, we vary the meteoritic bombardment rate from $3 \times 10^{23}$ $gGy^{-1}$ to $1 \times 10^{25}$ $gGy^{-1}$ in equal increments over 15 simulations. We follow the work of K. J. Zahnle et al. (2020) and assume impactors of $10^{22}$ g that would produce 0.65 bar of molecular hydrogen upon impact. Consequently, the surface influx of $H_2$ varies from $8.56 \times 10^9$ to $2.85 \times 10^{11}$ $cm^{-2}$ $s^{-1}$.

#### 2.2.3. Multiparameter Study

Informed by the effects of the individual parameters, described later, we choose to study the coupled effects of C/O ratio, semimajor axis, and host star. We combine the methods described above to construct 2925 models (15 C/O ratios and different semimajor axes for each of the 13 stars).

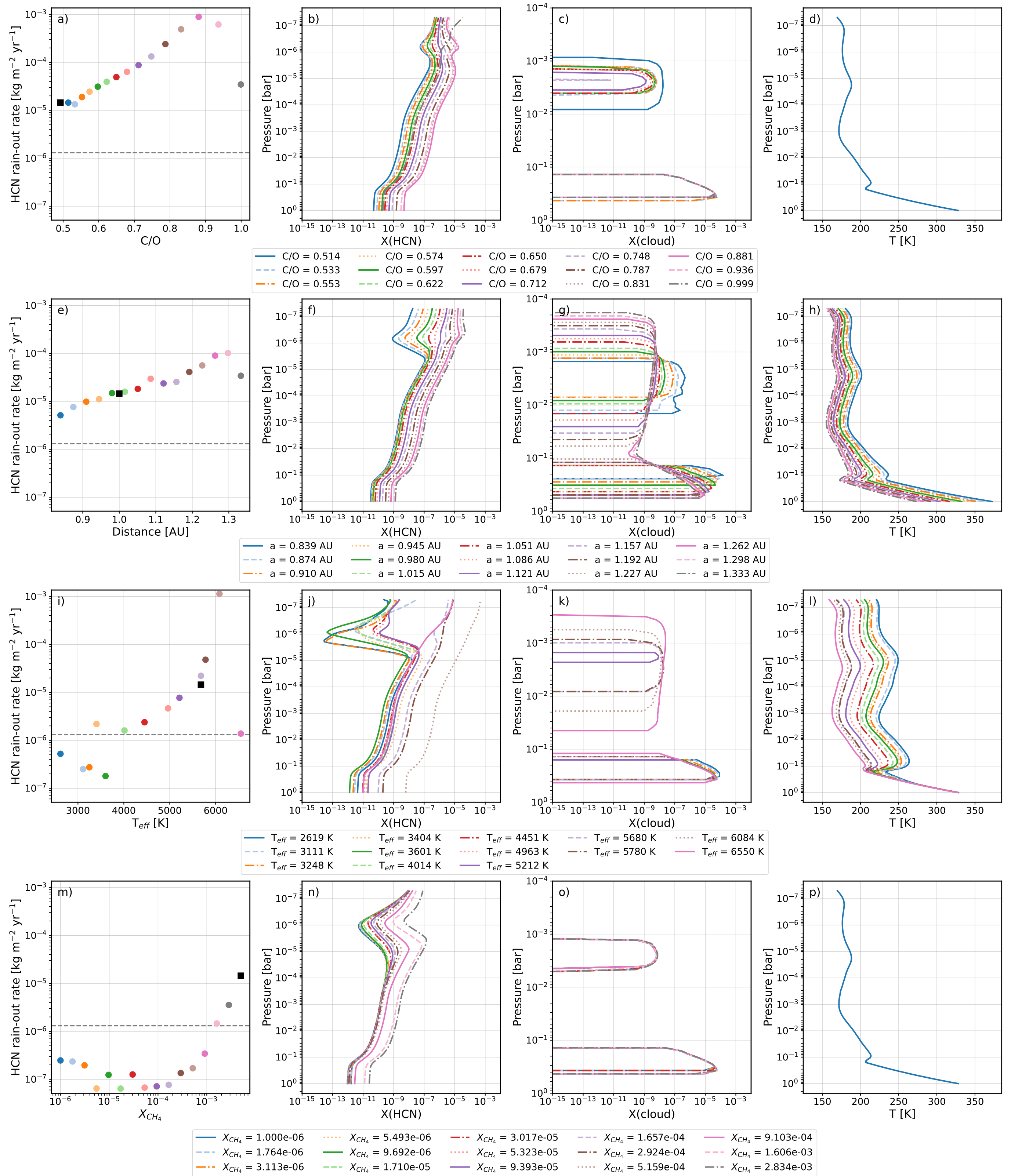

**Figure 2.** HCN rainout rates (first column), vertical profiles of HCN and water condensates (second and third column) and *T*–*P* profiles (last column, calculated using HELIOS; M. Malik et al. 2017, 2019a, 2019b; E. A. Whittaker et al. 2022) for four explored physical parameters (C/O, semimajor axis, type for host star and initial $CH_4$ mixing ratios, respectively from top to bottom row). HCN rainout rate from our baseline Archean simulation is marked with a black square, and the gray dashed line represents our upper estimate for exogenous HCN delivery.

The sensitivity of HCN to initial $CH_4$ levels suggests complex interactions with other atmospheric parameters that may amplify or dampen these production pathways. However, given the current lack of tight constraints on exoplanetary methane, we maintain the 5000 ppm baseline derived from the geological data provided by D. C. Catling & K. J. Zahnle

**Table 2**
Summary of Stellar and Orbital Parameters

| Name | $T_{eff}$ (K) | $\log(L/L_{\odot})$ | $R$ ($R_{\odot}$) | Distance (pc) | Type | $a$ (au) | Range of $a$ (au) | Range of $S_{eff}$ ($S_{\odot}$) | References |
|---|---|---|---|---|---|---|---|---|---|
| TRAPPIST-1[a] | 2619 | −3.26 | 0.1192 | 12.467 | M8V | 0.0269 | 0.0231–0.0350 | 1.0299–0.4486 | (1, 2) |
| GJ1214[b] | 3111 | −2.41 | 0.2162 | 14.642 | M4.5V | 0.0697 | 0.0610–0.0918 | 1.0455–0.4617 | (1, 2) |
| GJ729[a] | 3248 | −2.4 | 0.2 | 2.976 | M4V | 0.0719 | 0.061–0.094 | 1.0699–0.4506 | (1, 2) |
| GJ674[a] | 3404 | −1.77 | 0.3646 | 4.553 | M3V | 0.1451 | 0.124–0.192 | 1.1045–0.4607 | (1, 2) |
| GJ15A[a] | 3601 | −1.655 | 0.3754 | 3.562 | M2V | 0.1679 | 0.142–0.223 | 1.0975–0.4450 | (1, 2) |
| GJ676A[a] | 4014 | −1.079 | 0.6488 | 15.98 | M0V | 0.3653 | 0.308–0.485 | 0.8788–0.3544 | (1, 2) |
| HD85512[b] | 4451 | −0.778 | 0.6924 | 11.277 | K6V | 0.4808 | 0.408–0.642 | 1.0016–0.4045 | (1, 2) |
| HD40307[b] | 4963 | −0.585 | 0.7167 | 12.932 | K2.5V | 0.6228 | 0.523–0.823 | 0.9506–0.3839 | (1, 2) |
| HD97658[b] | 5212 | −0.455 | 0.728 | 21.563 | K1V | 0.6960 | 0.588–0.920 | 1.0145–0.4144 | (1, 2) |
| Early Sun[c] | 5680 | −0.128 | 0.892 | … | G2V | 1 | 0.839–1.333 | 1.0580–0.4191 | (3) |
| Sun[d] | 5780 | 0 | 1 | … | G2V | 1.1643 | 0.99–1.55 | 1.0203–0.4162 | (2) |
| HD149026[e] | 6084 | 0.42 | 1.4604 | 75.8643 | G0V | 1.8887 | 1.591–2.513 | 1.0391–0.4165 | (2) |
| WASP17[e] | 6550 | 0.613 | 1.5732 | 405.908 | F4 | 2.3367 | 1.960–3.098 | 1.0678–0.4274 | (2) |

**Notes.** Stellar parameters are the effective temperature ($T_{eff}$), the luminosity ($L$), and radius ($R$) in solar units ($L_{\odot}$ and $R_{\odot}$), the distance, and the type of the star. Orbital parameters are the semimajor axis ($a$), its range, and the effective solar radiation ($S_{eff}$). Spectra sources.

**References.** (1) A. Brown et al. (2023), P. R. Behr et al. (2023); (2) J. L. Christiansen et al. (2025); (3) M. W. Claire et al. (2012), J. N. Bahcall et al. (2001).

[a] D. J. Wilson et al. (2021).

[b] K. France et al. (2016), A. Youngblood et al. (2016), R. O. P. Loyd et al. (2016).

[c] J. N. Bahcall et al. (2001).

[d] C. A. Gueymard (2018).

[e] P. R. Behr et al. (2023).

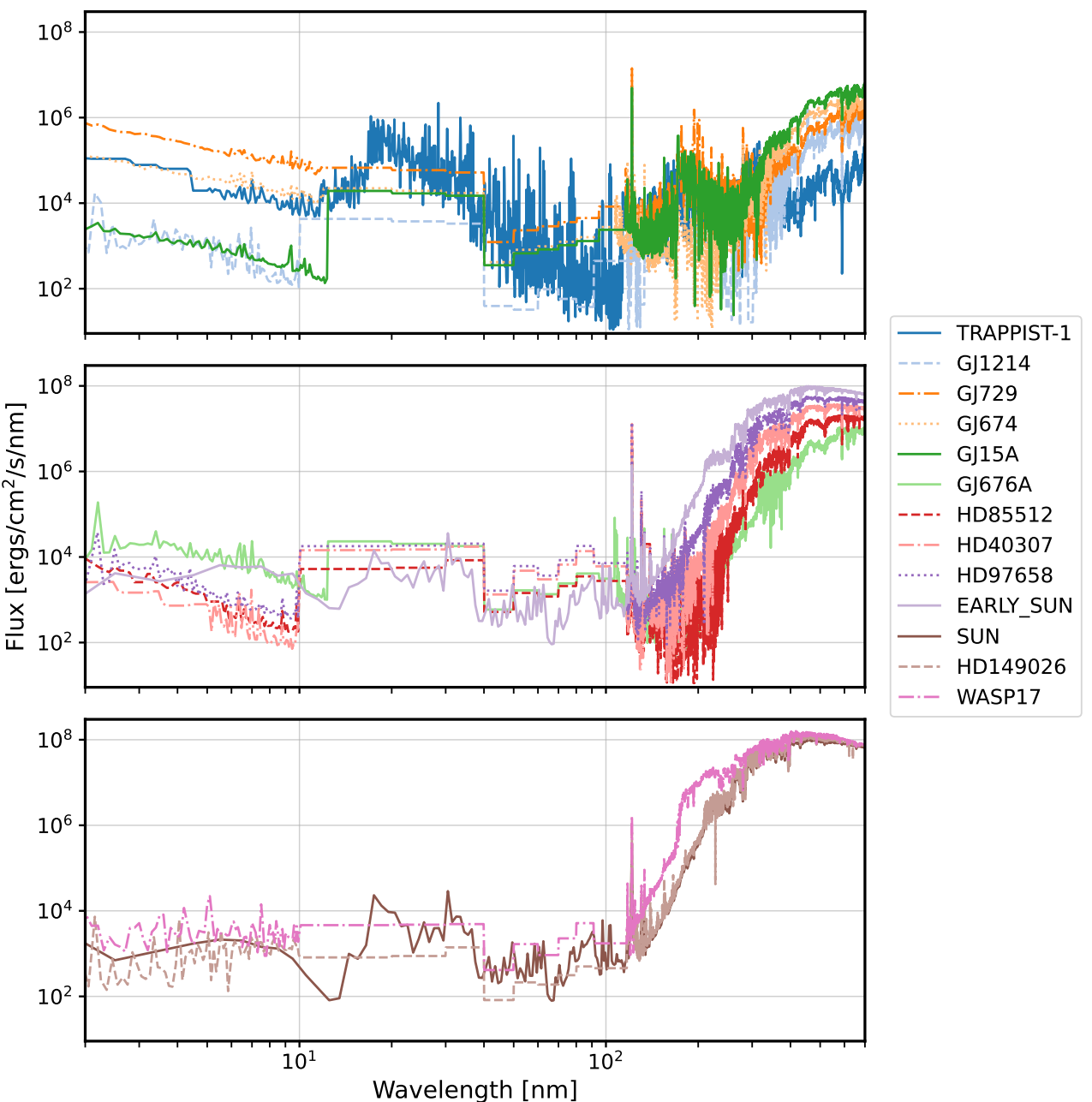

**Figure 3.** All 13 stellar spectra used in our study scaled to the respective stellar surface.

(2020). We set up the limits of the semimajor axes for each star so that the surface temperatures would range from 275 ± 1.5 K to 370.7 ± 1.5 K when calculated with the same atmospheric composition as the Archean Earth. The range of semimajor axes varies substantially by stellar host, so we convert the semimajor axes into effective stellar radiation ($S_{eff}$ in $S_{\oplus}$) received by our model planets using Equation (3) of R. K. Kopparapu et al. (2013):

$$S_{eff} = \frac{L}{L_{\odot}}\left(\frac{1\,\mathrm{au}}{d}\right)^2, \tag{3}$$

where $d$ is the semimajor axis (Astronomical unit), $L$ is the luminosity of the star and $L_{\odot}$ is the current solar luminosity. This means that $S_{eff} = 1$ for today's Earth. The conversion provides us with an easier and more intuitive way to understand how the semimajor axis controls HCN chemistry. Table 2 shows the semimajor axis $S_{eff}$ ranges.

## 3. Results

We report a HCN rainout rate of $1.446 \times 10^{-5}$ kg m$^{-2}$ yr$^{-1}$ for our baseline Archean Earth model (black square on Figure 2(a)). Compared to the works of B. K. D. Pearce et al. (2022), our model resulted in a higher HCN vertical mixing ratio profile and orders of magnitude more HCN rained out onto the planetary surface. When compared to F. Tian et al. (2011), our model produced less HCN, but rained out a similar amount, suggesting a more efficient rainout method. The main differences between our and previous studies ( F. Tian et al. 2011; B. K. D. Pearce et al. 2022) are that we parameterize the HCN rainout based on the vertical flux of water vapor (Section 2.1) instead of assuming a constant value, and that we explore a more diverse set of rocky exoplanets.

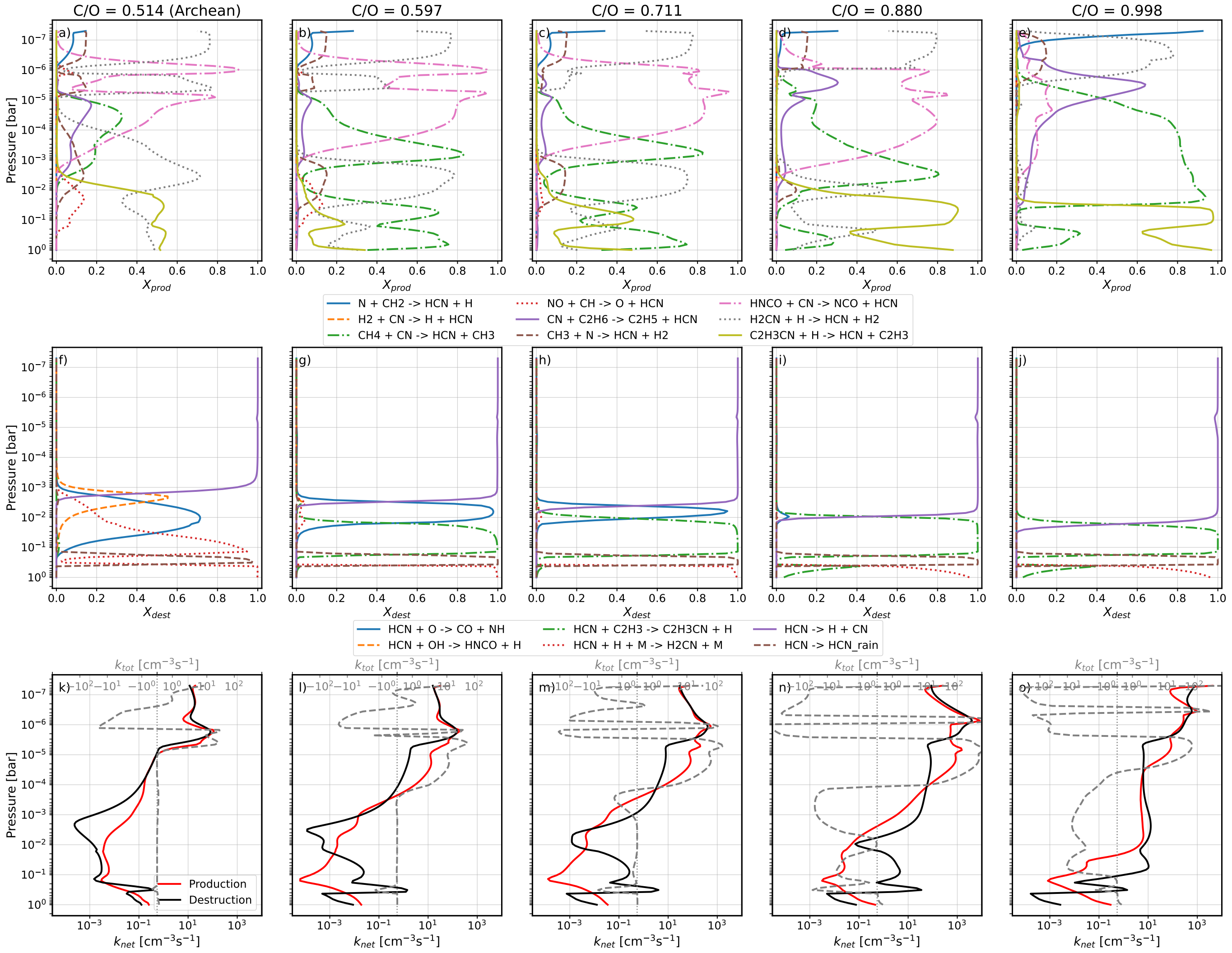

**Figure 4.** Important reactions and their relative contribution to HCN production (first row) and destruction (second row), along with the net production and destruction rates and their difference, the total reaction rate of HCN (third row) for chosen C/O ratios.

### 3.1. Influence of the C/O Ratio

We expect HCN mixing ratios and rainout rates to increase with increasing C/O ratios due to the enhanced production and subsequent atmospheric chemistry of HCN in more reducing environments (P. B. Rimmer & S. Rugheimer 2019; D. C. Catling & K. J. Zahnle 2020). This expectation is met by our models (Figures 2(a), (b)), although at C/O $\gtrsim 0.9$, the HCN rainout rate and the corresponding HCN mixing ratio profile decrease, except at the highest altitudes ($P \lesssim 10^{-6}$ bar). Our baseline Archean Earth simulation uses the least reducing atmosphere, so HCN rainout due to C/O is at a minimum. The lower cloud deck (Figure 2(c)), from where the rainfall originates, is insensitive to changes in C/O, suggesting the increase in HCN and its rainout rate is due to a chemical rather than a purely physical process.

The net production and loss rates of HCN, along with the total rate (production minus loss), shown by Figure 4(k)–(o), reveal that the amount of HCN available in the atmosphere is mainly driven by its reactions in the middle of the atmosphere ($10^{-6} \lesssim P \lesssim 10^{-2}$ bar). The total reaction rate for HCN is positive in the upper-middle atmosphere and increases with C/O ratio for C/O $\lesssim 0.9$ before decreasing again, while the destruction of HCN consistently strengthens in the lower-middle atmosphere. The latter takes effect at higher C/O ratios and is generally weaker than the former, which results in the initial increase in the HCN mixing ratio and rainout rate. However, for C/O $\gtrsim 0.9$, loss of HCN dominates a greater part of the middle atmosphere over HCN production, resulting in the decreased HCN mixing ratios and rainout rates.

We compare the relative importance of the most important production and loss reactions of HCN. Figures 4(a)–(e) show that the chemical production of HCN at lower pressures ($P \lesssim 10^{-3}$ bar) is driven by $H_2CN$+H→HCN+$H_2$ and HNCO+CN→NCO+HCN for most C/O ratios. For the Archean Earth case, i.e., lowest C/O ratio, HNCO+CN peaks at $P \simeq 10^{-6}$ and $P \simeq 10^{-5}$ bar, while $H_2CN$+H drives HCN production at other pressures. At the point where the C/O ratio reaches and exceeds 0.88, HNCO+CN becomes the dominant HCN production reaction for $P \gtrsim 10^{-6}$ bar; however, it completely diminishes for the highest C/O ratio and is replaced by CN+$C_2H_6$→$C_2H_5$+HCN and $CH_4$+CN→HCN+$CH_3$. At higher pressures ($P \gtrsim 10^{-3}$ bar), $H_2CN$+H and $C_2H_3CN$+H→HCN+$C_2H_3$ drive HCN production for most

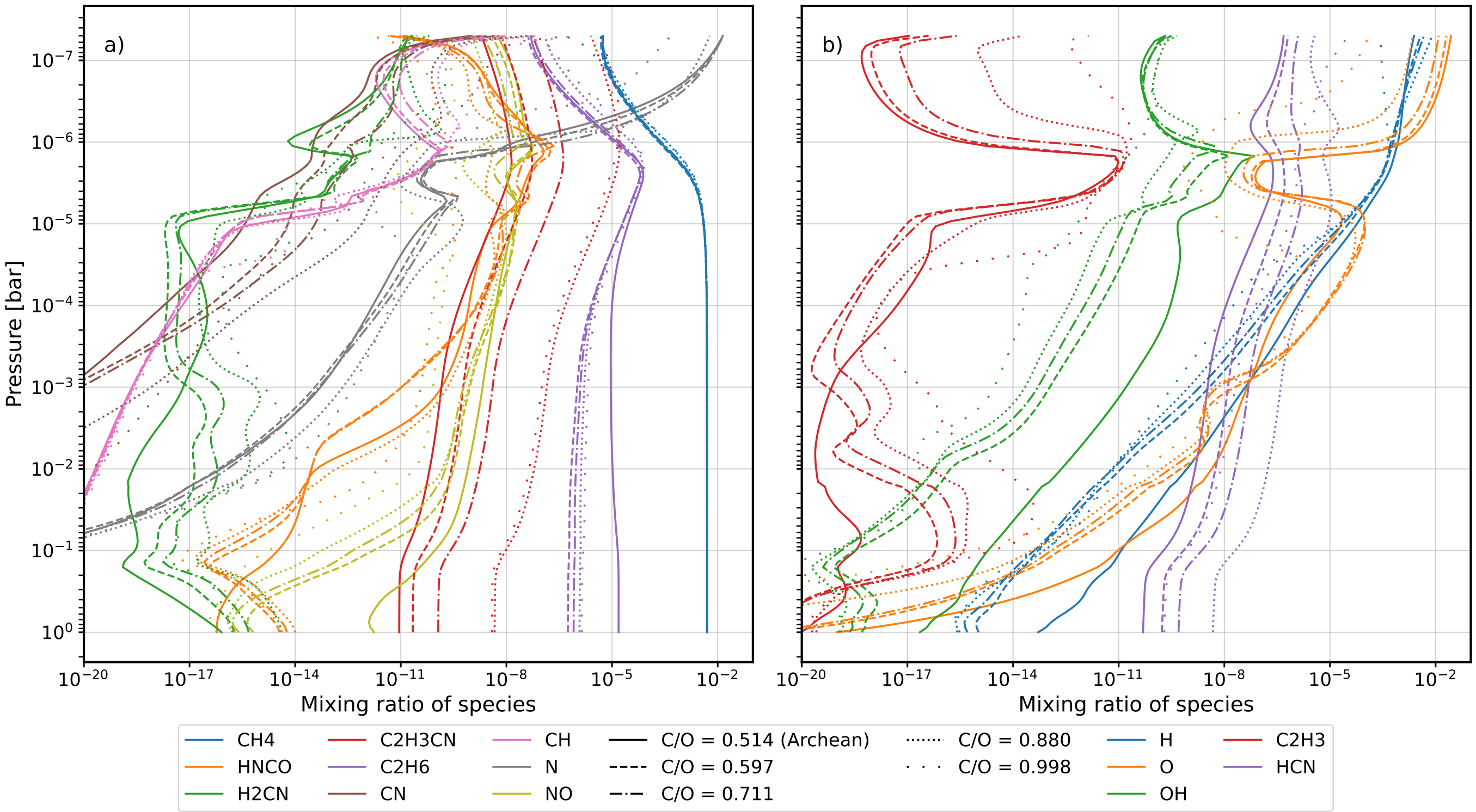

**Figure 5.** Vertical structure of various species contributing to HCN production or destruction for chosen C/O ratios.

C/O ratios. For intermediate and the highest C/O ratios, $CH_4$+CN also contributes to the formation of HCN. At most pressures, the change in the relative importance of reactions could be traced by the change in the mixing ratios of the reactant (Figure 5), except for the highest pressures. This suggests that chemical pathways are more complex and more sensitive to the C/O ratio at high pressures.

Figures 4(f)–(j) show that HCN loss is driven by photodissociation at $P \lesssim 10^{-3}$ bar for the Archean Earth case, and that this region expands toward higher pressures with C/O ratio, reaching $P \simeq 10^{-2}$ bar for the highest C/O ratio. At higher pressures and for the lowest C/O ratio, HCN+O→CO+NH, HCN+OH→HNCO+H, the three-body reaction (HCN+H+M→$H_2$CN+M, with M being the third body), and rainout are the most important loss reactions of HCN. The latter two reactions dominate the highest pressures and are constant over C/O ratios. On the other hand, with increasing C/O ratio, HCN+OH quickly diminishes, while HCN+O gradually weakens and is pushed toward lower pressures until it diminishes by C/O=0.88. Simultaneously, HCN+$C_2H_3$ → $C_2H_3$CN+H becomes the dominant HCN loss reaction between the photodissociation region and the cloud deck. These changes coincide with the change in the vertical profiles of the reactants (Figure 5).

### 3.2. *Influence of the Semimajor Axis*

Increasing the semimajor axis beyond what we assumed for our baseline Archean Earth increases the HCN rainout rate and mixing ratio (Figures 2(e), (f)). The increase in the HCN rainout rate is linear, but rainout rates decrease significantly at the highest values of the semimajor axis due to cold temperatures, which reduces wet scavenging efficiency and rain formation. The upper atmospheric HCN mixing ratio trough ($\sim 10^{-6}$ bar, see Figure 2(f)) becomes weaker as photodissociation of HCN becomes less important with higher values of the semimajor axis, eventually resulting in a small HCN peak. Increasing the semimajor axis affects the vertical structure of HCN and the cloud deck (Figures 2(f), (g)), suggesting substantive roles for chemical and physical processes in rainout rate changes.

We discuss the net and total reaction rates of HCN along with the relative importance of reactions here, but we place the relevant figures in Appendix B for brevity. As the net production and loss rates of HCN increase differently with semimajor axes, the total reaction rate strengthens in the middle of the atmosphere (Figures 9(k)-(o)). This phenomenon starts at lower pressures and expands toward higher pressures, turning the total rate of HCN positive in the entirety of the middle atmosphere by the highest semimajor axis, resulting in more HCN in this and the lower part of the atmosphere. This is accompanied with a thickening lower cloud deck at $P \gtrsim 10^{-1}$ bar, increasing HCN rainout rate with semimajor axis. The slight decrease at the highest values of the semimajor axis is mainly due to the reduced amount of water rain forming and its reduced ability for wet scavenging, which is caused by low temperatures.

Figures 9(a)–(e) shows that HCN production is again driven by $H_2$CN+H and HNCO+CN at lower pressures ($P \lesssim 10^{-3}$ bar) for all semimajor axes, with HNCO+CN only having two peaks at $P \simeq 10^{-6}$ and $P \simeq 10^{-5}$ bar for the closest orbits but dominating most of this part of the atmosphere for the furthest orbits. On planets with the smallest semimajor axis, CN+$C_2H_6$→$C_2H_5$+HCN and $CH_4$+CN→HCN+$CH_3$ also contribute to HCN production at $10^{-5} \lesssim P \lesssim 10^{-3}$ bar, which is likely due to an increase in CN and $C_2H_6$ mixing ratios (Figure 12) thanks to the strong irradiation. At higher pressures ($P \gtrsim 10^{-3}$ bar), $H_2$CN+H and $C_2H_3$CN+H are again the main sources of HCN for all semimajor axes. The $H_2$CN mixing

ratio in this part of the atmosphere first slightly decreases, then increases with the semimajor axis, while $C_2H_3CN$ increases monotonically and is always ∼6 orders of magnitude higher (Figure 12). The change in the relative importance of these two reactions follows that of the $H_2CN$ mixing ratio, suggesting that $H_2CN$ forms HCN more efficiently than $C_2H_3CN$ at these pressures.

HCN destruction (Figures 9(f)–(j)) is driven by photodissociation at $P \lesssim 10^{-3}$ bar for all semimajor axes. With increasing semimajor axis, however, the peak of photodissociation rate shifts toward lower pressures and strengthens (Figures 9(k)–(o)). This is likely due to the weakening radiation received by the planet that cannot penetrate its atmosphere as much, which also results in higher HCN abundances that in turn increase the photodissociation rate. At higher pressures ($P \gtrsim 10^{-3}$ bar), HCN+O, HCN+OH, HCN+H+M, and rainout drive the loss of HCN. The first two reactions mainly destroy HCN at the lower pressure end of this region, with HCN+O dominating at the closest orbits and HCN+OH dominating at the furthest orbits, which is in alignment with the changes in the vertical profiles of the reactant (Figure 12). The latter two reactions are important at the highest pressures, and their importance is relatively constant over semimajor axes, although rain reduces for the furthest orbits due to cold temperatures.

### *3.3. Influence of Stellar Irradiation*

Stellar and orbital configurations used in this study (Table 2) reveal a generally increasing HCN rainout rate with effective stellar temperature ($T_{\rm eff}$), as shown in Figure 2(i). Our baseline Archean Earth simulation rains out more HCN than most of the model planets that are irradiated by other host stars. The spectra used (Figure 3) affect the tropopause temperature for the cooler stars, which in turn changes the cloud structure (Figures 2(k), (l)). For these calculations, however, the HCN vertical structure is affected more than the lower cloud deck (Figures 2(j), (k)), suggesting that chemical processes are more likely to be responsible for driving the change in HCN rainout rate observed. Stronger short-wavelength ($\lambda \leqslant 200$ nm) radiation of M-type stars ($T_{\rm eff} \leqslant 4014$ K) leads to a significant trough in HCN mixing ratio in the upper atmosphere. This substructure is gradually smoothed with increasing $T_{\rm eff}$, leading to higher mixing ratios followed by enhanced transport rates onto the planetary surface, with the exception of the hottest star used in our study. We find similar behavior for the mixing ratios of hydrocarbons, such as $CH_4$, $C_2H_3$, and $C_2H_6$, in the upper atmosphere (Figure 13). We find noncontinuous behavior in the rainout rate and mixing ratios of HCN in the case of M-dwarfs that we believe is due to the spectra used, which have substantial variations for $\lambda < 400$ nm (Figure 3).

The net loss and production rates of HCN (Figure 10(k)–(o)) show that different stellar irradiation penetrates to and peaks at different depths in the atmosphere. With increasing effective temperature, HCN destruction strengthens in the upper atmosphere while production is generally higher than destruction in the middle atmosphere, resulting in an increase of the HCN rainout rate and mixing ratio. The latter is not the case for the hottest and only F star, WASP-17 ($T_{\rm eff} = 6550$ K), for which there is only a relatively small positive total reaction rate in the middle atmosphere, which is not enough to drive the HCN mixing ratio against the strong loss in the upper atmosphere. This results in the drop of the HCN rainout rate and mixing ratio observed.

HCN production (Figure 10(a)–(e)) at lower pressures ($P \lesssim 10^{-3}$ bar) is driven by photochemical radicals for planets orbiting M stars via N+$CH_2$ → HCN+H and NO+CH→O +HCN, although $H_2CN$+H dominates toward higher pressures. With increasing $T_{\rm eff}$, $H_2CN$+H and HNCO+CN gradually become the main drivers of HCN production, with HNCO+CN starting off, again, at $P \simeq 10^{-6}$ and $P \simeq 10^{-5}$ bar, and becoming the main source of HCN by the hottest G star (HD149026, $T_{\rm eff} = 6084$ K). For the hottest star, WASP-17 ($T_{\rm eff} = 6550$ K), HNCO+H is replaced by CH4+CN and H2CN+H toward higher pressures. These changes coincide with the changes in the mixing ratios of the reactants (Figure 13). At higher pressures ($P \gtrsim 10^{-3}$ bar), $H_2CN$+H and $C_2H_3CN$+H are the main sources of HCN for all stellar types, except for the coldest and hottest stars, for which the latter reaction does not significantly contribute to HCN production. This is due to the low mixing ratios of $C_2H_3CN$ for these two cases (Figure 13). As before, $H_2CN$+H appears to be a more efficient source of HCN as $H_2CN$ mixing ratios are consistently lower than those of $C_2H_3CN$ in this region.

HCN loss (Figure 10(f)–(j)) at $P \lesssim 10^{-3}$ bar is driven by photodissociation for all stellar types, though the pressure limit of this region slightly increases, and the peak in the photodissociation rate shifts upward and broadens with $T_{\rm eff}$ (Figure 10(k)–(o)). Furthermore, the weakest photodissociation is observed for the two K stars (HD85512 and HD40307, with $T_{\rm eff} = 4451$ and 4963 K), which have the lowest flux in the ∼100–200 nm wavelength range (Figure 3). We find that the mixing ratios of hydrocarbons, such as $C_2H_3$ and $C_2H_6$, are enhanced with increasing stellar effective temperature (Figure 13), and their shielding effect is likely responsible for the shift of the photodissociation peak toward higher altitudes. At higher pressures ($P \gtrsim 10^{-3}$ bar), HCN+O, HCN +OH, HCN+H+M, and rainout drive the loss of HCN for most stellar host types. For the lowest $T_{\rm eff}$, HCN+OH has a negligible contribution to HCN loss, but its importance gradually increases and equals that of HCN+O by the hottest stellar hosts. The K stars mentioned earlier (HD85512 and HD40307) do not fit this pattern as HCN+$C_2H_3$ dominates over HCN+O and HCN+OH. The pattern and the deviation of the K stars from it coincide with the changes in the mixing ratios of the reactants (Figure 13). We further note that the pressure range of rainout of HCN gradually shrinks with increasing stellar effective temperature, which is compensated for by the increasing HCN abundances, resulting in more HCN rainout. For the hottest star, WASP-17 ($T_{\rm eff} = 6550$ K), there is not enough HCN in the atmosphere that leads to the falling HCN rainout rate noted before.

### *3.4. Influence of the Initial $CH_4$ Concentration*

We find that the HCN vertical profile and rainout rate exhibit a nonlinear relationship toward the initial CH4 mixing ratio (Figures 2(m), (n)). Their values slightly decrease from the lowest $CH_4$ abundance (1 ppm) to ∼10 ppm (rainout rates drop by a factor of ∼2), after which HCN mixing ratios increase while HCN rainout rate stagnates until ∼500 ppm of methane, after which both values increase sharply. The minimum HCN rainout rate is achieved for 10-500 ppm of initial methane, and it is around $6 \times 10^{-8}$ kg m$^{-2}$ yr$^{-1}$, which is more than two orders of magnitude lower than that of the baseline Archean Earth. The cloud structure (Figure 2(o)) does not vary with initial $CH_4$ concentration, suggesting that the difference is driven by changes in atmospheric chemistry. The relative flatness, then sharp increase in HCN rainout rate with increasing availability of atmospheric $CH_4$, suggests that there is a threshold above which a new chemical pathway opens for HCN production.

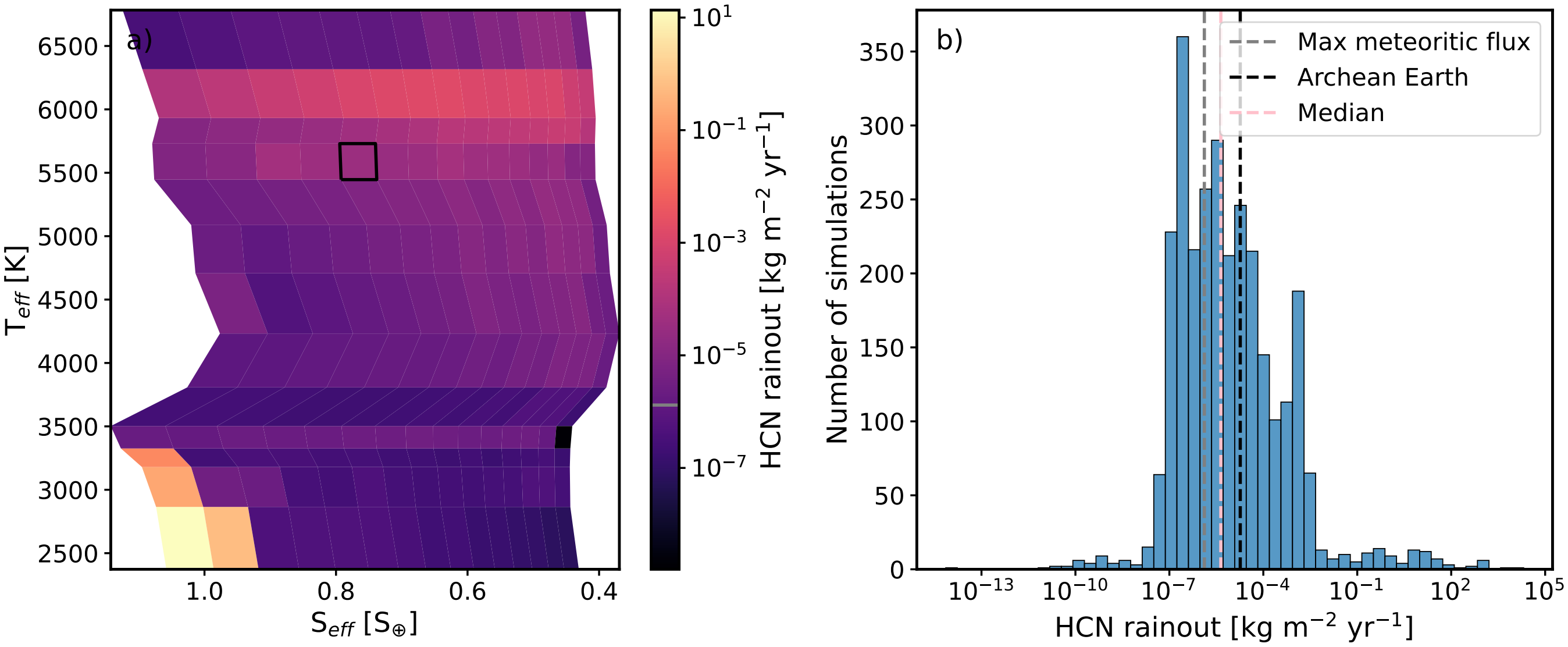

**Figure 6.** HCN rainout rate over the conjoint parameter space of stellar effective temperature ($T_{\mathrm{eff}}$), effective solar flux in current Earth values ($S_{\mathrm{eff}}$), and C/O ratio. (a) Baseline Archean Earth C/O ratio of 0.514 with the black box on the plot and gray horizontal line on the colorbar representing the Archean Earth model and estimated maximum HCN delivery rate by meteorites, respectively; (b) Histogram of all simulated HCN rainout rates with gray, black, and pink vertical dashed lines representing the estimated maximum exogenous HCN delivery rate, the Archean Earth model, and the median HCN rainout rate, respectively.

The net loss and production rates of HCN (Figure 11(k)–(o)) result in a total reaction rate that is close to zero for the lowest initial $CH_4$ abundances, which results in the low HCN rainout rates observed. Starting with ~500 ppm $CH_4$, the total reaction rate becomes positive in the upper-middle atmosphere with increasing strength and pressure range as the initial $CH_4$ increases. This results in more HCN in the atmosphere and so higher HCN rainout rates.

HCN production (Figure 11(a)–(e)) at lower pressures ($P \lesssim 10^{-3}$ bar) and at low initial $CH_4$ abundances is driven by the reaction HNCO+CN. As $CH_4$ reaches ~500 ppm, the $H_2CN$ + H pathway becomes the dominant and more efficient driver. This shift is marked by a positive net reaction rate for HCN (peaking at $10^{-3}$, $10^{-5}$, and $10^{-7}$ bar), correlating with a sharp increase in $H_2CN$ mixing ratios (Figure 14). At higher pressures ($P \gtrsim 10^{-3}$ bar), HCN production begins with NO +CH and $H_2CN$+H. At intermediate initial $CH_4$ abundances, the $CH_4$+CN reaction emerges as a contributor. At the highest methane concentrations, production is dominated by $H_2CN$+H and $C_2H_3CN$+H, as available methane preferentially forms $H_2CN$ and $C_2H_3CN$ intermediates rather than HCN directly.

The loss of HCN (Figure 11(f)–(j)) is driven almost exclusively by photodissociation at lower pressures ($P \lesssim 10^{-3}$ bar). At higher pressures, the reaction HCN+O acts as a universal sink across the parameter space. Supplementary destruction occurs via HCN+$C_2H_3$ at intermediate, and through HCN+OH and three-body reactions at the highest initial CH4 concentrations, tracking with the mixing ratios of those specific reactants (Figure 14).

### 3.5. *Influence of Meteoritic Bombardment Rate*

We find that varying the meteoritic bombardment rate does not significantly change the HCN rainout rate and vertical structure (Figure 15). This is unexpected, which we argue is due to the nature of the mean-state 1D model. Meteoritic impacts will most likely reduce and heat the atmosphere on a local scale, but the 1D model by definition describes the hemispheric mean state (for more details, we refer our reader to the original VULCAN papers S.-M. Tsai et al. 2017, 2021). Consequently, the vertical profiles of HCN and water condensates vary little, resulting in insignificant differences in HCN rainout rates across simulations.

### 3.6. *Influence of Multiple Parameters*

We construct a conjoint parameter space by varying the C/O ratio, semimajor axis, and stellar host type of our model planets while keeping them in their host star's HZ. First, we consider a small part of this 3D parameter space by fixing the C/O ratio to 0.514, corresponding to our Archean Earth, and exploring the influence of stellar effective temperatures and semimajor axes (Table 2). To ease comparison, we convert the latter into effective stellar irradiation ($S_{\mathrm{eff}}$), which is unity for today's Earth. The $1.446 \times 10^{-5}$ kg m$^{-2}$ yr$^{-1}$ HCN rainout rate of our baseline simulation is greater than ~72% (140 out of 195) of the simulations reported here. We find that the general trends seen in the single-parameter studies are robust (Figure 6(a)). As such, the higher HCN rainout rates compared to our baseline Archean Earth are achieved on model planets orbiting G stars, especially toward the outer edge of their respective HZ, and orbiting the hottest G star. We also find that model planets in the outer half of the HZ of K stars also yield HCN rainout rates comparable to those of the Archean Earth. Contrary to the individual parameter tendencies, there are three M stars, TRAPPIST-1, GJ1214, and GJ729, for which the closest orbits yield the HCN rainout rates that are not only comparable but surpass those of planets orbiting around G stars. The highest HCN rainout rate of 13.58 kg m$^{-2}$ yr$^{-1}$ is achieved on the model planet on the closest orbit around the M star TRAPPIST-1 at semimajor axis $a = 0.0231$ au.

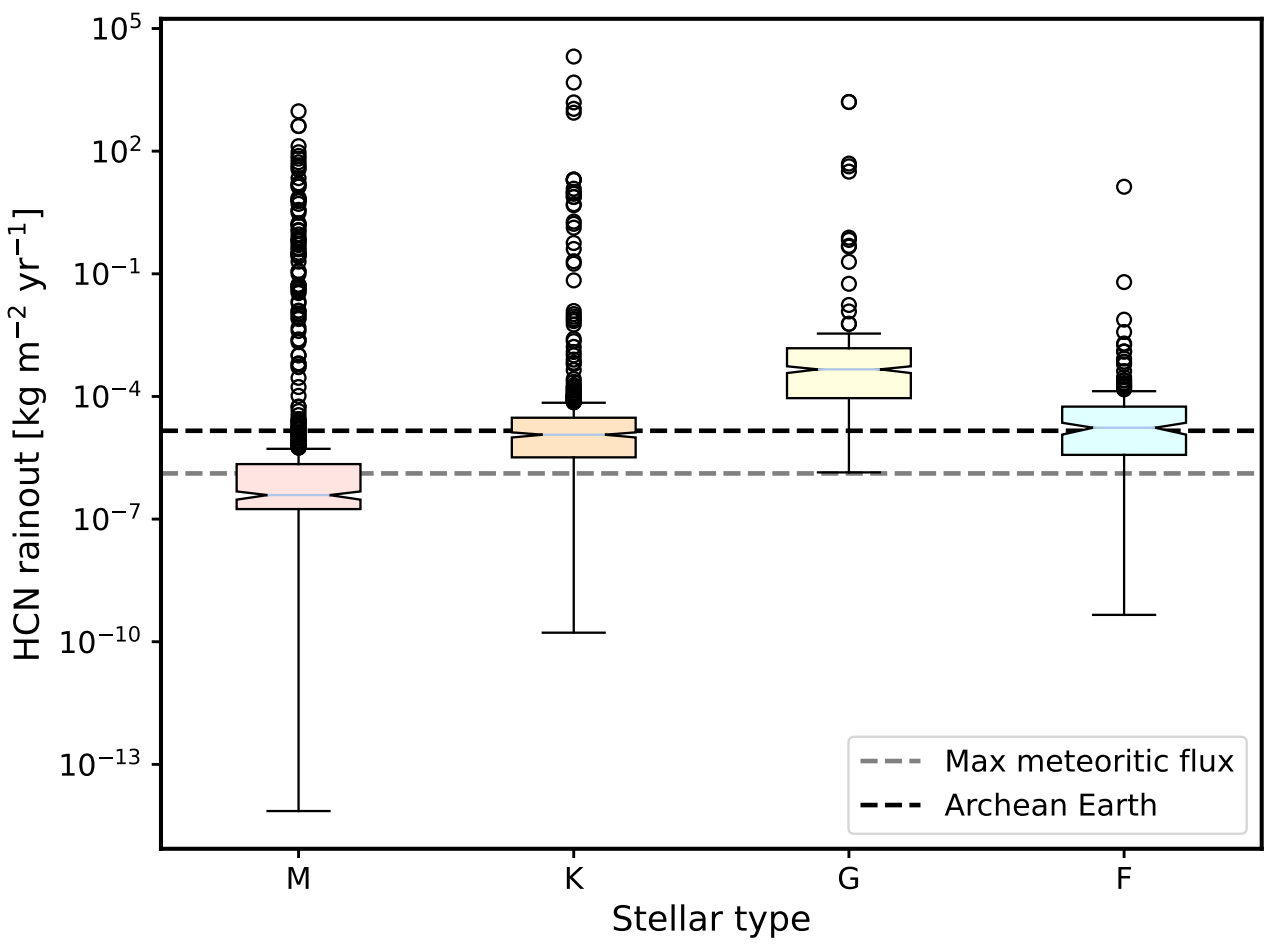

**Figure 7.** Box plot representing HCN rainout rates in our conjoint parameter studies grouped by stellar type. The circles are the outlying data points, the notches on the box represent the 95% confidence interval around the median, and the dashed horizontal lines in gray and black represent the estimated maximum exogenous HCN delivery rate and the rainout rate of our Archean Earth model, respectively.

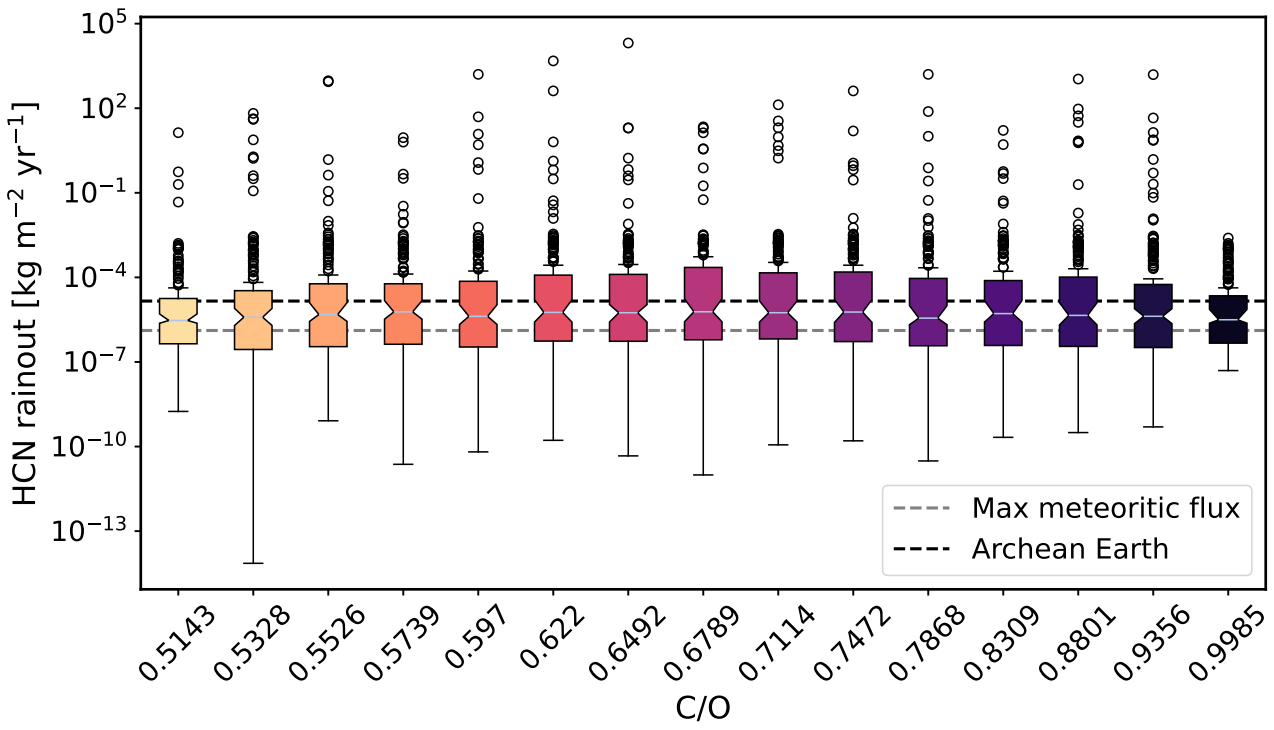

**Figure 8.** Same as Figure 7 but for C/O ratios.

When considering the entirety of our parameter space, an interesting picture emerges. Around 41% of the 2925 model planets produced a higher HCN rainout rate than our baseline Archean Earth model, with most HCN rainout rates within two orders of magnitude of this value (Figure 6(b)). Generally, model planets orbiting G stars produce the highest HCN rainout rates, followed by planets around K and F stars, while planets with an M host star have the lowest median HCN rainout rate (Figure 7), which is consistent with our individual parameter study. We find that the largest variances of rainout rates are for the coolest stars, and this variance decreases with stellar effective temperature. The median HCN rainout rate for each C/O ratio is within a factor of 2, although with different variances (Figure 8). This insensitivity to C/O ratio is contrary to our previous findings and suggests that the competing roles of atmospheric physics and chemistry are such that the physical environment dictates HCN chemistry more than the innate reducing capabilities of the atmosphere within the boundaries of this study. Guided by our individual parameter studies, we suggest that one of the main deciding factors of HCN rainout rates is how the two main, competing production pathways of HCN, $H_2CN$+H and HNCO+CN, are affected by both the physical and chemical environment. Our conjoint results here may highlight how these two pathways are more sensitive to the physical environment than the chemical, at least for a given initial $CH_4$ abundance. Our calculations suggest that lying close to either edge of the HZ increases the probability of producing larger HCN rainout rates, irrespective of the host star or C/O ratio.

HCN could be delivered to the planetary surface by impacting carbonaceous chondrite meteorites in the form of HCN or cyanide, of which the latter would form HCN when dissolved into water (S. Pizzarello 2012; K. E. Smith et al. 2019). Therefore, we compare the HCN rainout rates of our study with the possible HCN delivery by such meteorites. We use the cyanide concentration for the Murchison meteorite, 400 nanomole g$^{-1}$ (S. Pizzarello 2012). We note that this value differs from the 95 nanomole g$^{-1}$ from a previous study (K. E. Smith et al. 2019), but the different methodologies and samples used likely explain this difference. Regardless, we use the higher value of 400 nanomole g$^{-1}$ and the highest mass delivery rate used in our experiments ($10^{25}$ g Gyr$^{-1}$) to construct our upper estimate of the global meteoritic influx of HCN, which we represent as a gray line on the colorbar in Figure 6(a) and as dashed gray lines in Figures 6(b), 7 and 8. We find that ~68% of our simulations result in more endogenous HCN delivery onto the planetary surface than our estimates for meteoritic delivery. We acknowledge that the local influence of meteorites would likely be more significant than implied by the global mean value, but nevertheless we find it significant that atmospheric chemistry by itself could deliver the dominant contribution of HCN to the planetary surface. This, we argue, suggests that one of the requirements for prebiotic chemistry to form the building blocks of life is likely more universal than previously thought.

## 4. Caveats

We use a 1D model that allows us to explore the parameter space needed to examine the robustness of our results, but it lacks details about 3D climate dynamics (e.g., D. E. Sergeev et al. 2022). Further work will explore these calculations using a 3D chemistry climate model, even with a reduced-order atmospheric chemistry network that is compatible and viable with such a model. We acknowledge that many of the planets in our study that orbit an M-dwarf star are likely to be tidally locked. However, we adopt the same diurnal behavior for all our planets to limit the number of parameters changed, and so we can study the effects of changing the C/O ratio, the semimajor axis, and/or the host star.

We also acknowledge the importance of assumed stellar spectra, especially in the 100–200 nm range, as the driver of prebiotic chemistry, especially photochemistry in the upper atmosphere (V. S. Airapetian et al. 2016; S. Ranjan et al. 2017; P. B. Rimmer & S. Rugheimer 2019). The resolution of stellar spectra and assumptions in their construction determine the photolysis rates and affect numerical stability in photochemical kinetics simulations. We smoothed a few spectra (Section 2.2.2) to avoid numerical instabilities, although we still experienced some along with noncontinuous results (not shown). On the geological timescales associated with the development of prebiotic chemistry, stellar evolution will change the spectrum (e.g., M. W. Claire et al. 2012) that subsequently will change the chemistry it can support. For example, time-varying UV can enhance photochemistry (T. Konings et al. 2022), whereas the inclusion of energetic particles can add new chemical pathways (V. S. Airapetian et al. 2016; K. Herbst et al. 2019). Consequently, accounting for stellar evolution, including any flaring activity, will help us to explore how planetary atmospheric chemistry may change and what implications that has for HCN delivery. For the purpose of this study and because of the nature of the steady-state photochemical model used, this approach is not feasible.

Furthermore, we acknowledge the limitations of our study on the effects of meteoritic bombardment. This method is not designed to capture local effects, e.g., heating of the atmosphere, the various chemical species formed upon and after the impact other than $H_2$, but rather considers global mean effects of impacts. We believe that studying the chemistry locally after impacts would be valuable; however, it is outside of the scope of this study.

We only examine a subset of planetary parameters that could potentially affect HCN chemistry, albeit the ones we consider to be the most important and the ones we argue that we can observationally constrain. We also assume Earth-like surface conditions, for we have no better prior knowledge about the available water and continental crust on rocky exoplanets. Such surface parameters would naturally dictate what further chemistry of HCN occurs after it is rained out from the atmosphere. Too much water would reduce the HCN concentration to insignificant values, or a lack of dry–wet phases would hinder the formation of the building blocks of life, as suggested in the WLP hypothesis. The emergence of life is not driven solely by the overall availability of HCN, and our aim is to study only the first step of the complicated biochemical chain of forming life. Our simulated HCN rainout rates span several orders of magnitude; however, to our knowledge, we do not know whether there exists a realistic environmental HCN concentration that could negatively affect the polymerization and subsequent reactions that are needed to form more complex molecules in the surface water.

## 5. Conclusions

HCN is a simple organic molecule that potentially plays a crucial role in the emergence of life on our planet via several different hypothesized reaction mechanisms in warm little ponds. In this study, we explore the importance of HCN delivery from atmospheric chemistry on Archean Earth compared to delivery from meteorites. We also examined the robustness of our results with respect to our assumptions about the C/O ratio of the atmosphere, the semimajor axis, the stellar type, and the initial methane budget. In doing so, we have mapped out a multiparameter space that is relevant to rocky exoplanets and the search for life outside our solar system.

For our calculations, we used a 1D steady-state photochemical model including our implementation of wet deposition to study the removal of HCN by rainout. Unlike previous studies, which approximated HCN rainout rates by gravitational settling (F. Tian et al. 2011; B. K. D. Pearce et al. 2022), we implement a time-independent, first-order approximation of water rain formation, coupled with Henry's law of solubility to estimate the delivery of HCN onto the planetary surface (see Section 2.1). We report a HCN rainout rate of $1.446 \times 10^{-5}$ kg m$^2$ yr$^{-1}$ for the Archean Earth, which is orders of magnitude higher than reported by previous studies (F. Tian et al. 2011; B. K. D. Pearce et al. 2022).

We find that the atmospheric HCN rainout rate is insensitive to the change in meteoritic bombardment rate, at least on a global scale. Varying the C/O ratio results in an increase of HCN production and rainout rates, as expected, with a more reducing environment capable of producing more organic material. However, the enhanced amount of (complex) organics for C/O ratios above ∼0.9 effectively destroys HCN mid-atmosphere, decreasing its mixing ratios and rainout rates. Studying the effects of the semimajor axis, we find that HCN is directly affected by the weakening photodissociation with increasing semimajor axis, which leads to higher HCN mixing ratios and rainout rates toward the outer edge of the habitable zone. At the furthest location, however, low temperatures reduce precipitation and wet removal efficiency, and, consequently, HCN rainout rate decreases. We also note a positive correlation between the effective temperature of the stellar host and HCN rainout rate with model planets orbiting G stars, e.g., the early Sun, yielding the highest HCN rainout rates. Here, this is due to changes in chemistry and cloud structure/precipitation. Model planets that orbit M stars produce the lowest HCN rainout rates, suggesting lower potential for prebiotic chemistry. S. Ranjan et al. (2017) also claimed lower potential for prebiotic chemistry around M stars, based on less radiation at UV wavelength, which is important for prebiotic chemistry at the surface, suggesting the possible importance of flares. We identify a critical $CH_4$ concentration threshold at approximately 500 ppm. Below this limit, HCN production is dominated by a single pathway involving HNCO, which is a likely byproduct of $CO_2$ chemistry. Above this threshold, a second, highly efficient pathway emerges via $H_2CN$, which originates from the $CH_4$ feedstock. The synergy of these dual pathways leads to a significant increase in total HCN abundance. Our results suggest that the specific availability of $CH_4$ is a primary driver of HCN chemistry. This adds nuance to earlier studies (e.g., P. B. Rimmer & S. Rugheimer 2019), which have emphasized the C/O ratio as the dominant factor over specific atmospheric species. We find that even with a fixed C/O ratio, the HCN yield is highly sensitive to the initial methane abundance, specifically above ∼500 ppm of $CH_4$.

From these individual parameter studies, we learn the importance of molecules such as $H_2CN$ and HNCO, which drive HCN chemistry over a wide range of the parameter space. We further note that HCN chemistry is also sensitive to CN and $C_3H_3$, even though these species often exhibit much lower mixing ratios than other species involved in HCN chemistry. We believe that this finding has an intriguing implication: there could be ubiquitous species that drive the HCN chemistry on a diverse range of rocky exoplanets.

Our conjoint study of C/O ratio, semimajor axis, and stellar host reveals that ∼68% of our model planets generate a larger HCN rainout rate than our estimate for global meteoritic rates, and ∼41% exceed our baseline Archean Earth model. This suggests that HCN chemistry and delivery onto the planetary surface are robust over a significant part of the parameter space studied. We argue that a critical implication of our finding is that the availability of HCN, and potentially other simple organics such as formaldehyde, is not the rate-limiting step for the prebiotic chemistry forming the building blocks of life. However, other factors not included in this study, such as surface conditions and topology, plate tectonics, climate stability, and stellar activity, could also play important roles, and therefore studying their effect on prebiotic chemistry is essential for our understanding of the emergence of life.

The most optimal Earth-like planets, from the point of view of HCN rainout rate, are planets orbiting M and G stars close to the inner and outer edge of their habitable zone, respectively. Although planets from the former part of the parameter space yield more HCN rain, the latter part of the parameter space is more voluminous, with many possible planets yielding HCN rainout rates that are comparable to or exceed those of the Archean Earth. Surprisingly, this is true regardless of the C/O studied, which illustrates the importance of the physical environment over the inherent reducing capabilities of the atmosphere for the production and rainout of HCN. Analytical and numerical calculations by R. J. Anslow et al. (2023) suggest that the highest likelihood of HCN survival by impacting comets is in the case of Earth-like planets orbiting Sun-like stars in a tightly packed system. This and our results could imply that Earth-like planets orbiting G stars could benefit from both endogenous and exogenous HCN production and delivery. On the other hand, surface prebiotic chemistry on planets in close-in orbits of M stars, experiencing high HCN rainout rates according to our models, would need to, and could, mainly rely on endogenous HCN production and delivery.

Our findings suggest that vital prebiotic molecules, such as HCN and its derivatives, could be ubiquitous and readily available on a wider range of rocky exoplanets than previously thought. Although the emergence of life is not directly implied by the availability of its building blocks, it could increase the probability of life existing beyond Earth, especially on planets orbiting G- and M-type stars. This conclusion could help direct future astronomical observations toward these specific types of planets to search for and characterize potentially habitable worlds.

## Acknowledgments

G.F. acknowledges his studentship by the Science and Technology Facilities Council [project reference: 2902875]. We are grateful to the Reviewer for their observations and suggestions that led to a stronger version of this study. We thank Ben Pearce for his openness in kindly sharing initial ideas about his own previous work on HCN and prebiotic chemistry, and ways he had overcome numerical issues that were similar to ours. We also thank Arturo Alberto Lira Barria and Greg Cooke for their help with overcoming numerical issues. Benjamin Benne is thanked for his helpful ideas in creating clear figures. Special thanks to Shang-Min Tsai and Matej Malik and their teams in creating and maintaining VULCAN and HELIOS, respectively, that made this work possible. This research has made use of the NASA Exoplanet Archive, which is operated by the California Institute of Technology, under contract with the National Aeronautics and Space Administration under the Exoplanet Exploration Program. This paper makes use of data from the first public release of the WASP data (Butters et al. 2010) as provided by the WASP consortium and services at the NASA Exoplanet Archive, which is operated by the California Institute of Technology, under contract with the National Aeronautics and Space Administration under the Exoplanet Exploration Program.

*Facility:* Exoplanet Archive (J. L. Christiansen et al. 2025; NASA Exoplanet Science Institute 2020).

*Software:* VULCAN (S.-M. Tsai et al. 2017, 2021), HELIOS (M. Malik et al. 2017, 2019a, 2019b; E. A. Whittaker et al. 2022).

## Author Contributions

G.F. and P.I.P. designed research, G.F. and M.B. performed research, G.F. analyzed data, and G.F., P.I.P., M.B., and K.R. wrote the paper.

## Appendix A
## 1D Model of Atmospheric Photochemistry

Here we describe the parts of the VULCAN model relevant to our study and refer the reader elsewhere for a broader model description (S.-M. Tsai et al. 2017, 2021). VULCAN solves mass continuity equations that describe the temporal ($t$) change of the number density of each species $i$ ($n_i$) in a given layer (or cell) of the atmosphere:

$$\frac{\partial n}{\partial t} = \mathcal{P}_i - \mathcal{L}_i - \frac{\partial \phi_i}{\partial z}, \tag{A1}$$

where $\mathcal{P}_i$ and $\mathcal{L}_i$ is the photochemical production and loss rates for the species $i$, respectively; $\phi_i$ is the corresponding transportation flux; and $z$ denotes the vertical height (S.-M. Tsai et al. 2017). The transport flux includes advection, eddy diffusion ($K_{zz}$), and molecular and thermal diffusion:

$$\phi_i = n_i v - K_{zz} n_{\text{total}} \frac{\partial X_i}{\partial z} - D_i\left(\frac{\partial n_i}{\partial z} - n_i\left(\frac{1}{H_i} + \frac{1+\alpha_T}{T}\frac{dT}{dz}\right)\right), \tag{A2}$$

where $v$ is the vertical wind velocity, $n_{\text{total}}$ is the total gas number density, $K_{zz}$ is the eddy-diffusion coefficient, $D_i$, $H_i$, and $X_i$ are the molecular diffusion coefficient, the scale height, and the mixing ratio of the $i$th species, while $\alpha_T$ is the thermal diffusion factor and $T$ is the temperature (S.-M. Tsai et al. 2021).

VULCAN permits the inclusion of atmospheric boundary conditions, such as volcanic degassing, ocean evaporation, or products of lightning chemistry at the surface, and atmospheric escape of H or $H_2$ at the top of the atmosphere. Many of these parameters are not well constrained and give a good opportunity to explore their effects on the steady-state atmospheric composition (R. Hu et al. 2012). Additionally, the model also provides flexibility with, for example, the chemical network and solar irradiation profile, allowing us to study various exoplanets in different stellar systems. The initial atmospheric composition is also customizable. As long as the C/O ratio is not altered and the initial molecules contain all atoms in the network, the steady-state solution will not depend on the choice of the starting species (S.-M. Tsai et al. 2017).

VULCAN comes with its own general chemical networks of which the most suitable for us is the N–C–H–O full photochemical network[7] that contains, at the time of writing, 418 forward thermochemical reactions—and the same number of reverse reactions of which VULCAN calculates the reaction rate coefficients using NASA polynomial thermochemical data (for more details on this process, see S.-M. Tsai et al. 2017)—and 57 photochemical reactions between more than 70 species. Although the choice of chemical network can affect the resulting steady-state solution (M. Braam et al. 2026), we choose the N-C-H-O full photochemical network for compatibility reasons.

[7] https://github.com/exoclime/VULCAN/blob/master/thermo/NCHO_full_photo_network.txt

# Appendix B
# Detailed Chemistry Figures for Individual Parameter Studies

Here we show the detailed chemical production and loss reactions, their contribution to HCN chemistry, and the total reaction rates of HCN for the individual parameter studies of semimajor axis, stellar host type, and initial methane budget (Figure 9, 10, 11). Furthermore, we show the vertical mixing ratios of more species involved in HCN chemistry for these studies (Figure 12, 13, 14). These figures guide the description of chemistry in Sections 3.2, 3.3, and 3.4. Last, we show the uniformity of HCN rainout rate with regards to meteoritic bombardment rate on Figure 15.

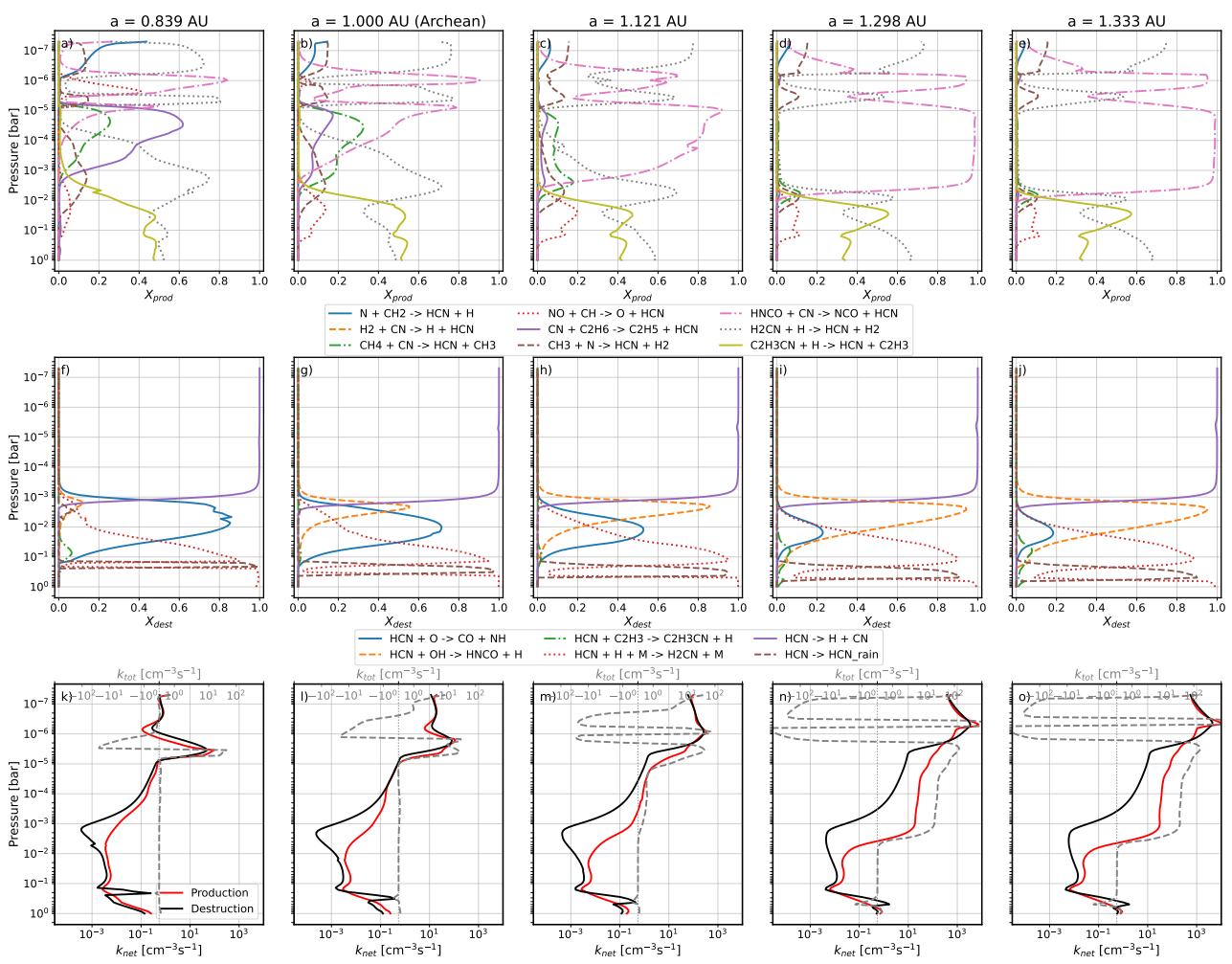

**Figure 9.** Same as Figure 4 but for chosen semimajor axes.

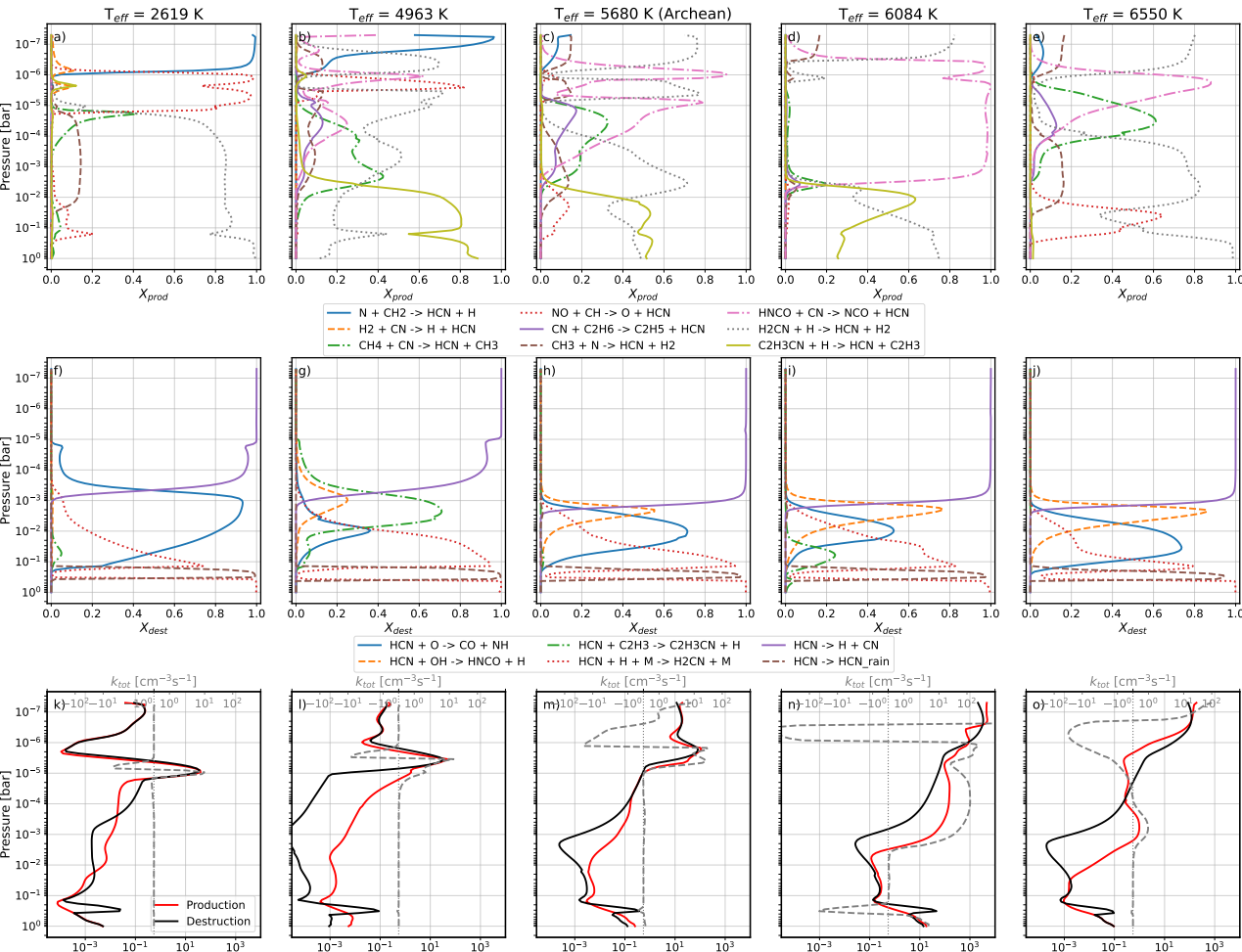

**Figure 10.** Same as Figure 4 but for chosen stellar hosts.

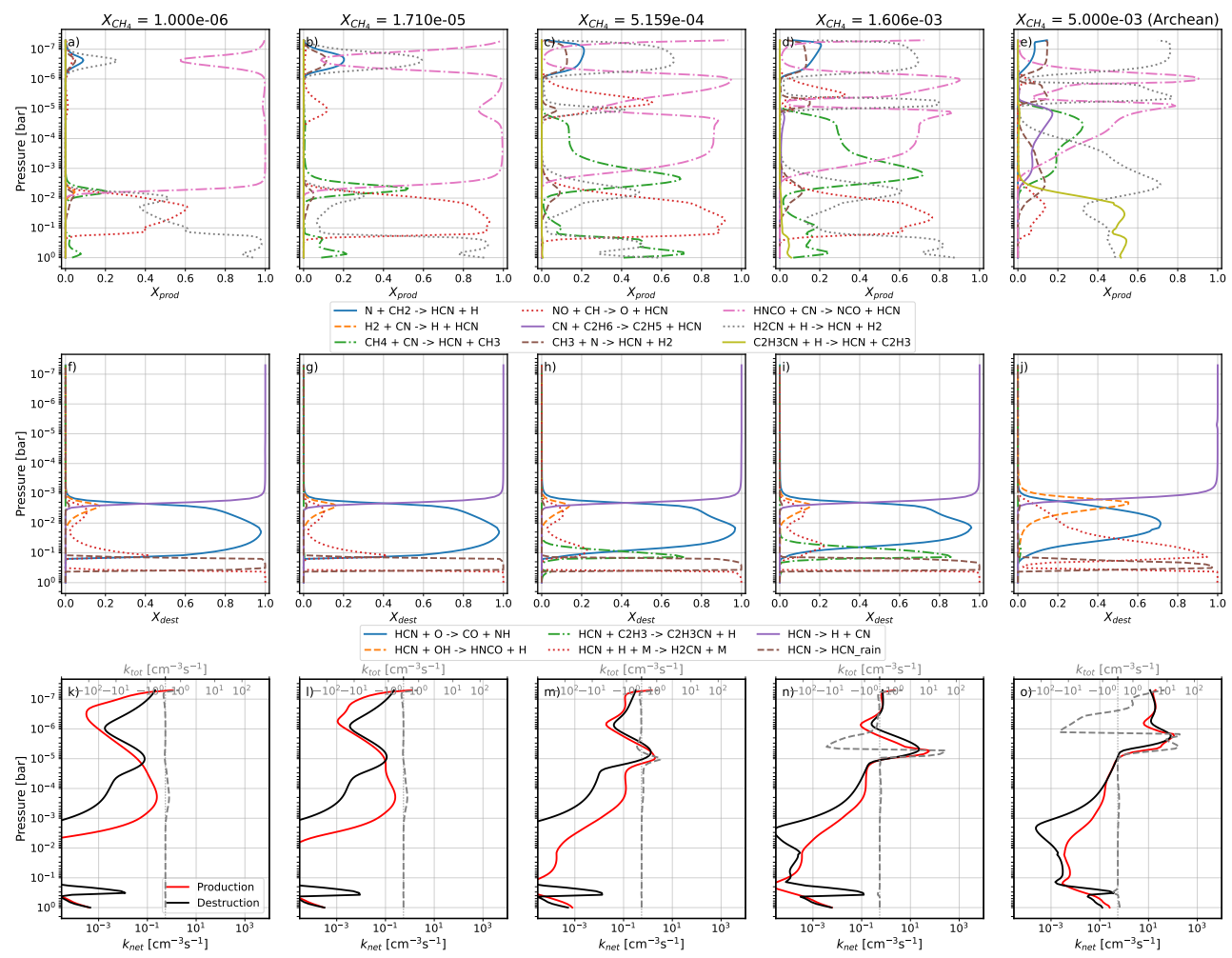

**Figure 11.** Same as Figure 4 but for initial $CH_4$ concentrations.

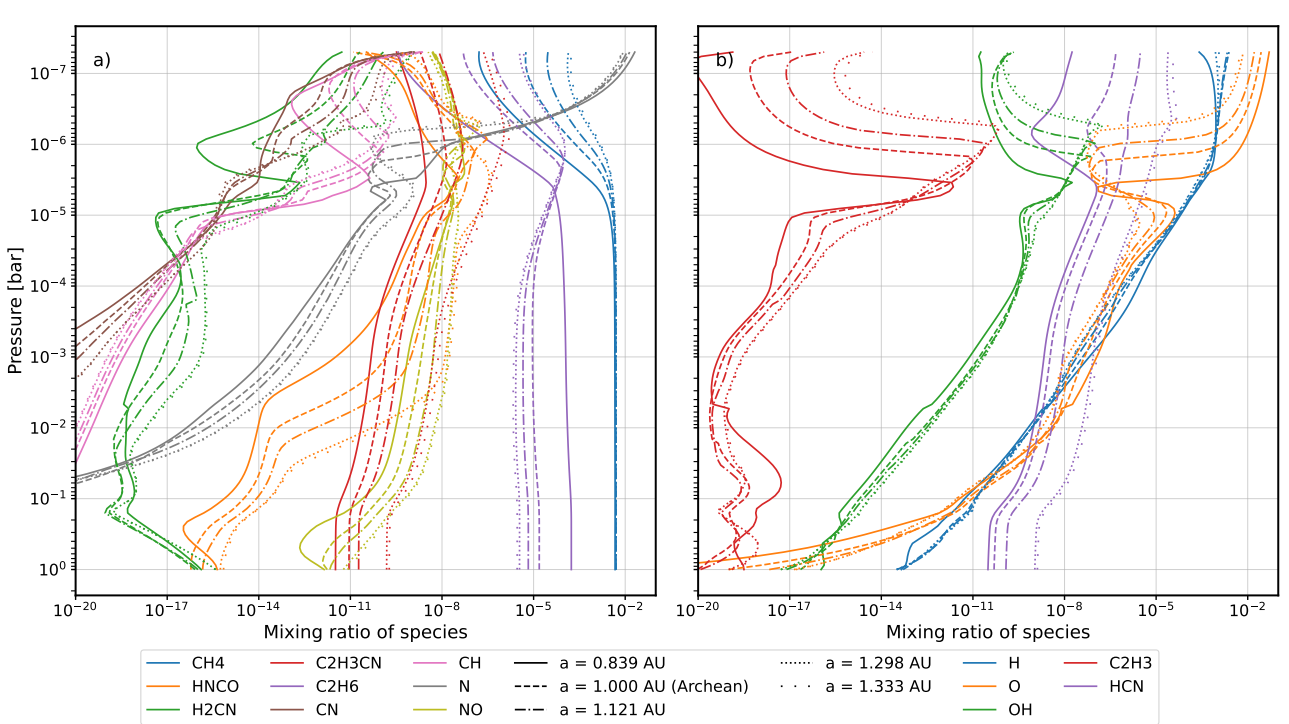

**Figure 12.** Same as Figure 5 but for chosen semimajor axes.

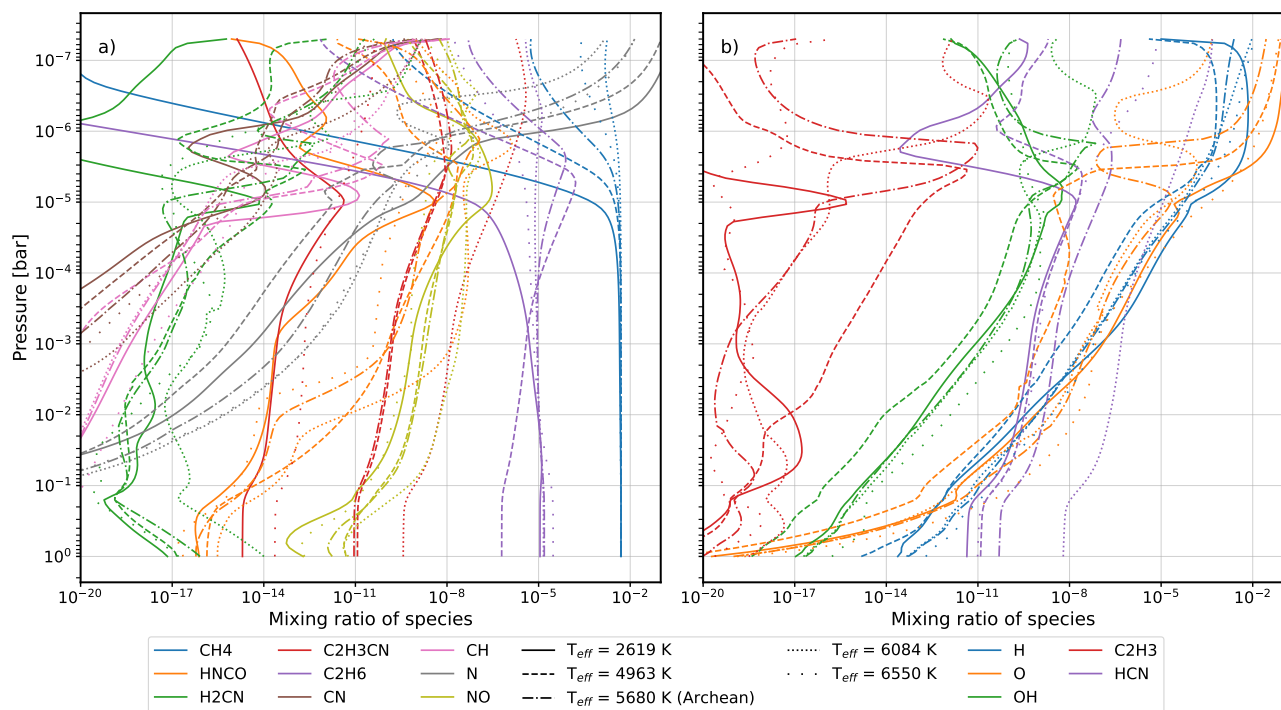

**Figure 13.** Same as Figure 5 but for chosen stellar hosts.

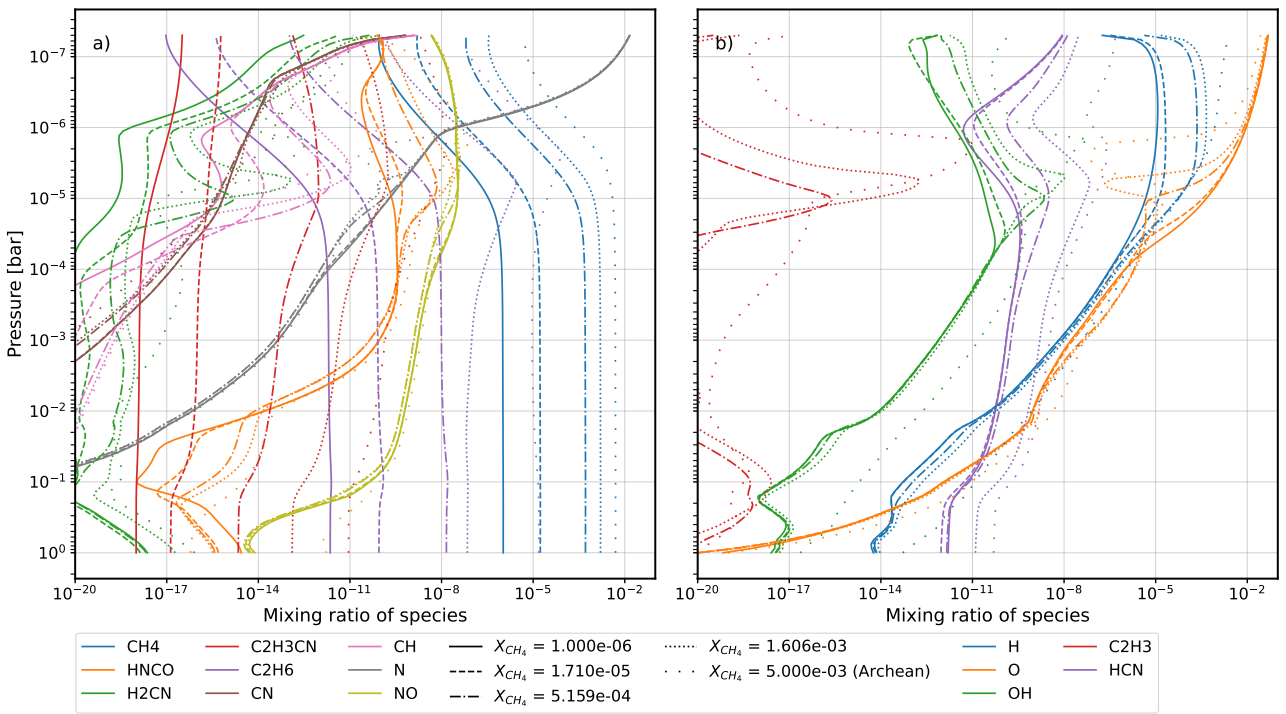

**Figure 14.** Same as Figure 5 but for initial $CH_4$ concentrations.

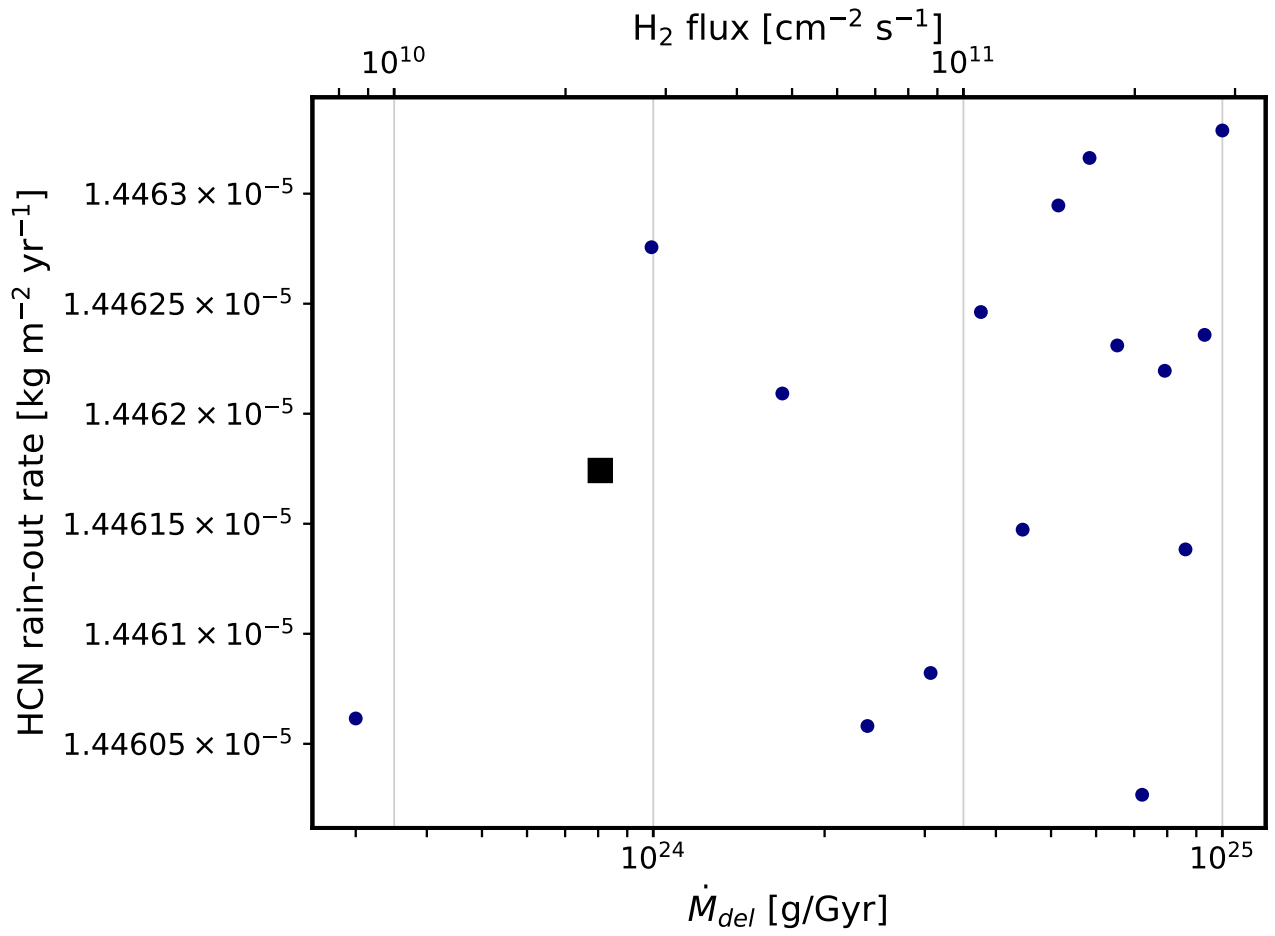

**Figure 15.** HCN rainout rates for varying meteoritic bombardment rates that change the surface boundary condition of $H_2$ flux.

## ORCID iDs

Gergely Friss https://orcid.org/0009-0008-5099-5698
Paul I. Palmer https://orcid.org/0000-0002-1487-0969
Marrick Braam https://orcid.org/0000-0002-9076-2361
Ken Rice https://orcid.org/0000-0002-6379-9185